\documentclass[aps,pra,superscriptaddress,twocolumn]{revtex4}
\usepackage{graphics}
\usepackage{amsmath}
\usepackage{graphicx}
\usepackage{amsfonts}
\usepackage{amssymb}
\begin{document}
\title{High-rate quantum cryptography in untrusted networks}

\begin{abstract}
We extend the field of continuous-variable quantum cryptography to
a network formulation where two honest parties connect to an
untrusted relay by insecure quantum links. To generate secret
correlations, they transmit coherent states to the relay where a
continuous-variable Bell detection is performed and the outcome
broadcast. Even though the detection could be fully corrupted and
the links subject to optimal coherent attacks, the honest parties
can still extract a secret key, achieving high rates when the
relay is proximal to one party, as typical in public networks with
access points or proxy servers. Our theory is confirmed by an
experiment generating key-rates which are orders of magnitude
higher than those achievable with discrete-variable protocols.
Thus, using the cheapest possible quantum resources, we
experimentally show the possibility of high-rate quantum key
distribution in network topologies where direct links are missing
between end-users and intermediate relays cannot be trusted.

\end{abstract}
\author{Stefano Pirandola}
\affiliation{Department of Computer Science, University of York,
York YO10 5GH, United Kingdom}
\author{Carlo Ottaviani}
\affiliation{Department of Computer Science, University of York,
York YO10 5GH, United Kingdom}
\author{Gaetana Spedalieri}
\affiliation{Department of Computer Science, University of York,
York YO10 5GH, United Kingdom}
\author{Christian Weedbrook}
\affiliation{Department of Physics, University of Toronto, Toronto
M5S 3G4, Canada and QKD Corp., 112 College St., Toronto M5G 1L6,
Canada}
\author{Samuel L. Braunstein}
\affiliation{Department of Computer Science, University of York,
York YO10 5GH, United Kingdom}
\author{Seth Lloyd}
\affiliation{MIT -- Department of Mechanical Engineering and
Research Laboratory of Electronics, Cambridge MA 02139, USA}
\author{Tobias Gehring}
\affiliation{Department of Physics, Technical University of
Denmark, Fysikvej, 2800 Kongens Lyngby, Denmark}
\author{Christian S. Jacobsen}
\affiliation{Department of Physics, Technical University of
Denmark, Fysikvej, 2800 Kongens Lyngby, Denmark}
\author{Ulrik L. Andersen}
\affiliation{Department of Physics, Technical University of
Denmark, Fysikvej, 2800 Kongens Lyngby, Denmark} \maketitle

Quantum key distribution (QKD)~\cite{Gisin,Scarani} is one of the
most active areas in quantum information~\cite{Wilde,RMP}, with a
number of in-field implementations, including the development of
metropolitan networks based on point-to-point QKD
protocols~\cite{SECOQC,SECOQC2,Tokyo1,Tokyo2}. A typical QKD
protocol involves two parties, conventionally called Alice and
Bob, who aim to generate a secret key by exchanging quantum
systems over an insecure communication channel. Security is
assessed against the most powerful attack on the channel, where an
eavesdropper, conventionally called Eve, perturbs the quantum
systems using the most general strategies allowed by quantum
mechanics.

While this theoretical analysis is fundamental for testing the
basic security of a protocol, it may be insufficient to guarantee
its viability in realistic implementations, where flaws in the
devices may provide alternative `side-channels'\ to be
attacked~\cite{SD4,SD6}. These weaknesses naturally arise in
realistic models of networks (e.g., the Internet) where two
end-users are not connected by direct lines but must exploit one
or more intermediate nodes, whose operation may be tampered with
by Eve.

In this scenario, Ref.~\cite{SidePRL} has recently\ introduced a
general method to guarantee security. Considering arbitrary
quantum systems as information carriers, Ref.~\cite{SidePRL}
designed a swapping-like protocol where secret correlations are
established by the measurement of a third untrusted party. This
idea of `measurement-device independence'\ has been independently
introduced in Ref.~\cite{Lo} in the specific context of qubits
(weak pulses and decoy states), with a series of further
investigations~\cite{Others,Oth2,Oth3,Oth4,Oth4b,Oth5,EXP1,EXP2,EXP3}.

In this paper we develop the notion of measurement-device
independence for bosonic systems. In this way we extend the field
of continuous-variable quantum
cryptography~\cite{RMP,Nicolas2001,Grangier,NoSwitch,TwowayPROTOCOL,Grangier2}
to a more robust network formulation. In fact, we consider the
basic network topology where Alice and Bob communicate by
connecting to an untrusted relay via insecure links. To create
secret correlations, they transmit random coherent states to the
relay where a continuous-variable Bell detection is performed and
the outcome broadcast. Despite the possibility that the relay
could be fully tampered with, and the links subject to optimal
coherent attacks, Alice and Bob are still able to extract a secret
key.

Our analysis shows that the optimal configuration of the protocol
corresponds to the relay being close to one party, e.g., Alice, in
which case remarkable rates can be achieved, orders of magnitude
higher than those achievable with
discrete-variable protocols over comparable distances~\cite{Lo,EXP1,EXP2,EXP3}%
. Our theoretical prediction is fully confirmed by a
proof-of-principle experiment, where $10^{-2}$ secret bits per
relay use are distributed at $10$dB loss in Bob's link, equivalent
to $50$km in standard optical fibre (at the loss rate of
$0.2$dB/km). Such rate corresponds to hundreds of kbits/s using
state-of-the-art clock rates at 75MHz~\cite{EXP3}. Furthermore,
assuming ideal reconciliation, our experiment shows a potential
rate of about $10^{-4}$ secret bits per relay use over a very
lossy link, with $34$dB loss corresponding to $170$km in fibre.

Note that this asymmetric configuration resembles the typical
topology of a public network where a user connects its device to a
proxy server to communicate with remote users. This setup is here
studied in the full optical regime but may also occur in a mixed
technology environment where a wireless device (e.g., the infrared
port of a laptop or phone) connects to a nearby access point,
which is the hub of a star network of remote users connected by
long optical fibres.

We remark that our formulation provides conceptual and practical
advantages with respect to traditional quantum cryptography. First
we move from a point-to-point to a robust end-to-end formulation,
which removes both trust and complexity from the middle nodes. In
particular, the fact that the relay is just performing a simple
detection on the incoming systems removes expensive quantum
sources from the network. This feature makes the key rate orders
of magnitude higher than that of protocols based on a central
distribution of entanglement~\cite{Ekert,WeedEPR}. Second we fully
exploit the advantages provided by continuous variables using the
cheapest possible quantum sources (coherent states) and the most
efficient quantum measurements (homodyne detectors). This other
feature makes our protocol easy to implement on realistic networks
with key rates out of the reach of any protocol based on
discrete-variable systems.

Thus, our work exploits the cheapest quantum resources to prove
high-rate QKD in untrusted networks where direct links are missing
between two end-users. This represents the first step towards the
realization of scalable models of secure quantum networks, based
on the end-to-end principle~\cite{endtoend} and the modern notion
of `reliability from unreliable parts'~\cite{Baran}.

\section{Results}

For simplicity, we start by describing the protocol in noiseless
links, explaining the basic mechanism of the relay. Then we
consider the most general eavesdropping strategy against the relay
and the links, showing how this strategy can be reduced to a
coherent Gaussian attack of the links only. Finally we derive the
secret-key rate of the protocol and we compare theoretical and
experimental performances in the optimal configuration.

\begin{figure}[ptbh] \vspace{-0.4cm}
\par
\begin{center}
\includegraphics[width=0.65\textwidth] {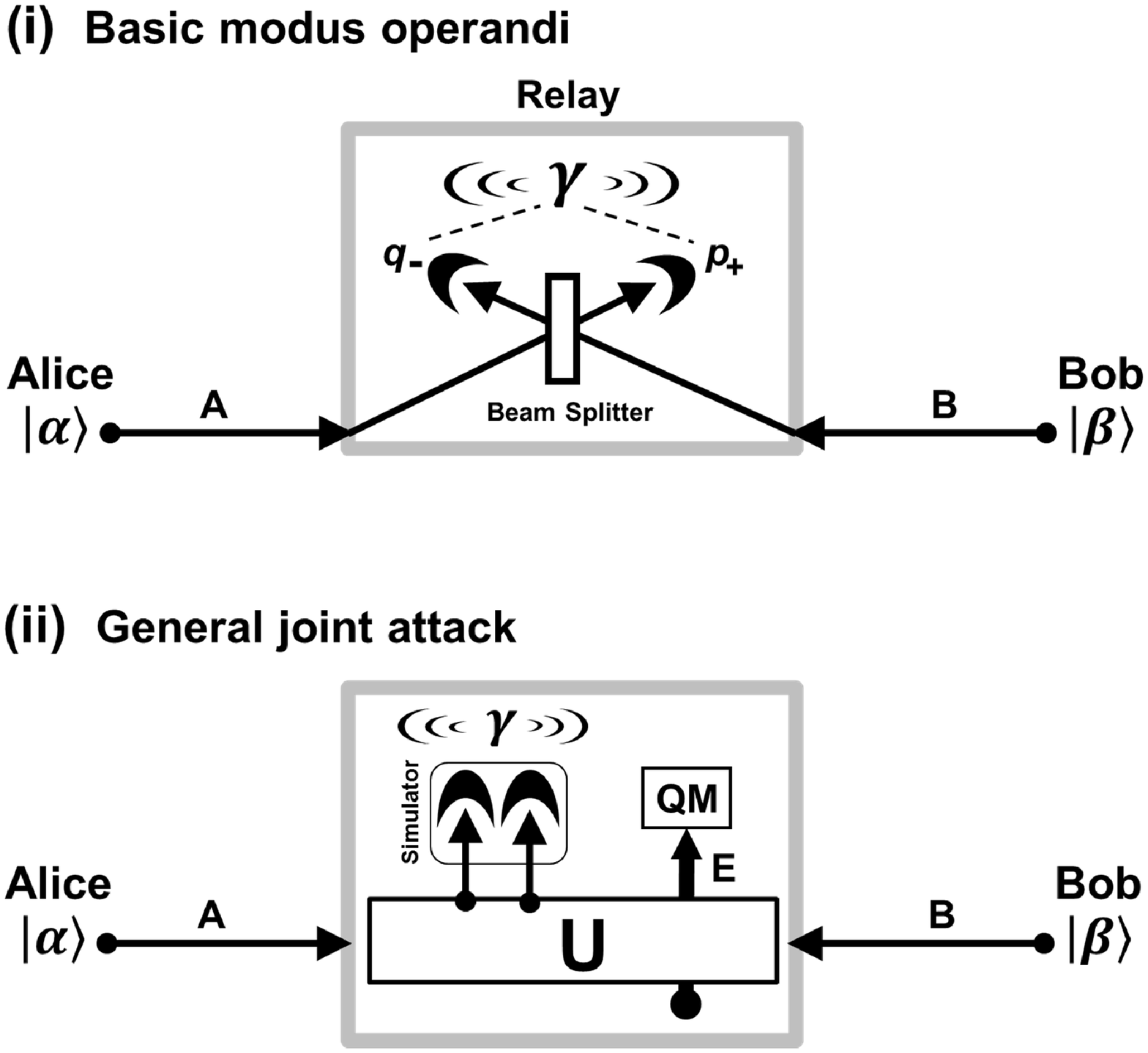}
\end{center}
\par
\vspace{-0.7cm}\caption{(i)~Modus operandi of the
continuous-variable Bell relay (see text for explanation).
(ii)~Joint attack of the protocol. In each use of the relay, modes
$A$ and $B$ unitarily interact with ancillary vacuum modes. Two
outputs simulate the relay, while the remaining outputs $E$ are
stored in a quantum memory (QM) measured by Eve at the end of the protocol.}%
\label{PMscheme}%
\end{figure}

\textbf{Basic idea.}--~Consider two distant parties, Alice and
Bob, aiming to share a secret key. At one side, Alice prepares a
mode $A$ in a coherent state $\left\vert \alpha\right\rangle $
whose amplitude $\alpha$ is modulated by a Gaussian distribution
with zero mean and large variance, equal to $\varphi \gg1$ in each
quadrature. At the other side, Bob prepares his mode $B$ in
another coherent state $\left\vert \beta\right\rangle $ whose
amplitude $\beta$ is modulated by the same Gaussian distribution
as Alice. Modes $A$ and $B$ are then sent to an intermediate
station, which is the continuous-variable Bell relay shown in
Fig.~\ref{PMscheme}(i). 

The relay performs a continuous-variable Bell detection on the
incoming modes, by mixing them in a balanced beam splitter whose
output ports are conjugately homodyned~\cite{BellFORMULA}. This
detection corresponds to measuring the quadrature operators
$\hat{q}_{-}=(\hat{q}_{A}-\hat{q}_{B})/\sqrt{2}$ and
$\hat{p}_{+}=(\hat{p}_{A}+\hat{p}_{B})/\sqrt{2}$, whose classical
outcomes are combined in a complex variable
$\gamma:=(q_{-}+ip_{+})/\sqrt{2}$ with probability $p(\gamma)$.
The outcome $\gamma$\ is then communicated to Alice and Bob via a
classical public channel.

In this process the relay acts as a correlator~\cite{SidePRL}. One
can check that the outcome $\gamma$ creates \textit{a-posteriori}
correlations between the parties, being equal to
$\gamma=\alpha-\beta^{\ast}+\hat{\delta}$ with $\hat{\delta}$
detection noise. As a result, the knowledge of $\gamma$ enables
each party to infer the variable of the other party by simple
postprocessing. For instance, Bob may compute
$\beta^{\ast}+\gamma=\alpha+\hat{\delta}$ decoding Alice's
variable~\cite{NOTA2way}. Thus, conditioned on $\gamma$, Alice and
Bob's mutual information increases from $I(\alpha:\beta)=0$ to
$I(\alpha:\beta|\gamma)>0$.

Averaging over all possible outputs $\gamma$, the honest parties
will share $I_{AB}=\int d^{2}\gamma
p(\gamma)I(\alpha:\beta|\gamma)$ mean bits per use of the relay,
which is logarithmically increasing in the modulation $\varphi$.
Despite Eve also having access to the classical communication and
operating the relay, she cannot steal any information, since she
only knows $\gamma$ and $I(\alpha:\gamma)=I(\beta:\gamma)=0$. As a
result, Eve is forced to attack the links and/or corrupt the
relay.

\textbf{Protocol under general eavesdropping}.--~The most general
eavesdropping strategy of our protocol is a joint attack involving
both the relay and the two links as depicted in
Fig.~\ref{PMscheme}(ii). In each use of the relay, Eve may
intercept the two modes, $A$ and $B$, and make them interact with
an ensemble of ancillary vacuum modes via a general unitary $U$.
Among the output modes, two are sent to a simulator of the relay,
where they are homodyned and the result $\gamma$ broadcast. The
remaining modes $E$ are stored in a quantum memory which is
measured at the end of the protocol.

Note that a more general attack may involve a unitary applied to
all modes which are transmitted over many uses of the relay.
However, this can always be reduced to the previous attack,
coherent within the single use, by assuming that Alice and Bob
perform random permutations on their data~\cite{Renner,Renner2}.
Also note that, in Eve's simulator, any other higher-rank
measurement of the quadratures can always be purified into a
rank-one measurement by enlarging the set of the ancillas $E$. If
other observables are measured or no detection occurs, the
communication of a fake variable $\gamma$ can be easily detected
from the analysis of the empirical data~\cite{SidePRL}.

In order to deal with the joint attack of Fig.~\ref{PMscheme}(ii),
Alice and Bob must retrieve the joint statistics of the variables
$\alpha$, $\beta$, and $\gamma$. Since the protocol is performed
many times, Alice and Bob can compare a small part of their data
via the public channel and reconstruct the probability
distribution $p(\alpha,\beta,\gamma)$. As we show in the
Supplementary Information, for any observed distribution $p(\alpha
,\beta,\gamma)$, the security of the protocol does not change if
we modify Eve's unitary $U$ in such a way that her simulator works
exactly as the original relay (so that the modes are mixed in a
balanced beam splitter and conjugately homodyned). Thus, we can
assume that the relay is properly operated (even if by Eve) with
the unitary $U$ restricted to be a coherent attack against the two
links.

The description of this attack can further be simplified. Since
the protocol is based on the Gaussian modulation and detection of
Gaussian states, its optimal eavesdropping is based on a Gaussian
unitary $U$~\cite{Raul}. Thus, from the first- and second-order
statistical moments of the observed distribution
$p(\alpha,\beta,\gamma)$, Alice and Bob construct a Gaussian
distribution $p_{G}(\alpha,\beta,\gamma)$ and design a
corresponding optimal Gaussian attack against the links.

From an operational point of view, the first-order moments are
used to construct the optimal estimators of each other variable,
while the second-order moments are used to derive the secret key
rate of the protocol. In particular, Alice and Bob are able to
compute their mutual information $I_{AB}$ and upperbound Eve's
stolen information $I_{E}$ via the Holevo bound. As long as the
condition $R:=I_{AB}-I_{E}>0$ is satisfied, they can postprocess
their data via standard procedures of error correction and privacy
amplification, and distill an average of $R$ secret bits per use
of the relay.

\begin{figure}[ptbh] \vspace{-1.6cm}
\par
\begin{center}
\includegraphics[width=0.55\textwidth] {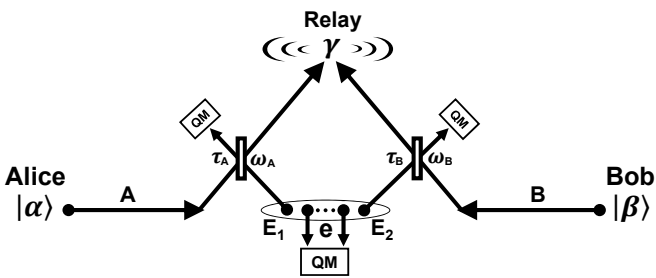}
\end{center}
\par
\vspace{-2.2cm}\caption{Protocol in the presence of a coherent
Gaussian attack. The two travelling modes, $A$ and $B$, are mixed
with two ancillary modes, $E_{1}$ and $E_{2}$, via two beam
splitters, with transmissivities $\tau_{A}$ and $\tau_{B}$, which
introduce thermal noise with variances, $\omega_{A}$ and
$\omega_{B}$, respectively. The ancillary modes belong to a
reservoir of ancillas ($E_{1},$~$\mathbf{e},~E_{2}$) which is
globally in a pure Gaussian state. All Eve's output is stored in a
quantum memory (QM)
measured at the end of the protocol.}%
\label{purification}%
\end{figure}

\textbf{Coherent Gaussian attack of the links}.--~Following the
previous reasoning, the cryptoanalysis of the protocol can be
reduced to studying a coherent Gaussian attack against the two
links, assuming a properly-working relay. Attacks of this kind can
be constructed by correlating two canonical
forms~\cite{Pirandola2009,canATTACKS}. The most realistic scenario
is the Gaussian attack depicted in
Fig.~\ref{purification}.

In this attack, the two travelling modes, $A$ and $B$, are mixed
with two ancillary modes, $E_{1}$ and $E_{2}$, by two beam
splitters with transmissivities $\tau_{A}$ and $\tau_{B}$,
respectively. These ancillary modes belong to a reservoir of
ancillas ($E_{1}$, $E_{2}$ plus an extra set $\mathbf{e}$) which
is globally described by a pure Gaussian state. The reduced state
$\sigma_{E_{1}E_{2}}$\ of the injected ancillas is a correlated
thermal state with zero mean and covariance matrix in the normal form%
\begin{equation}
\mathbf{V}_{E_{1}E_{2}}=\left(
\begin{array}
[c]{cc}%
\omega_{A}\mathbf{I} & \mathbf{G}\\
\mathbf{G} & \omega_{B}\mathbf{I}%
\end{array}
\right)  ,~\mathbf{G}:=\left(
\begin{array}
[c]{cc}%
g & 0\\
0 & g^{\prime}%
\end{array}
\right)  ,\label{EveCM_E1E2}%
\end{equation}
where $\omega_{A},\omega_{B}\geq1$ are the variances of the
thermal noise affecting each link, while $g$ and $g^{\prime}$ are
correlation parameters which must satisfy physical
constraints~\cite{NJP2013,TwomodePRA} (see Supplementary
Information). After interaction, all Eve's ancillas are stored in
a quantum memory, which is coherently measured at the end of the
protocol.

Note that our description of the attack is very general since any
two-mode Gaussian state can be transformed into the previous
normal form by local Gaussian unitaries~\cite{RMP}. In general,
the injected state $\sigma _{E_{1}E_{2}}$ can be separable or
entangled. The simplest case is when there are no correlations
($g=g^{\prime}=0$), so that $\sigma_{E_{1}E_{2}}$ is a tensor
product and the attack collapses into a collective attack with two
independent entangling cloners~\cite{Grangier}.

\textbf{Secret-key rate of the protocol}.--~In the attack of
Fig.~\ref{purification}, the relay provides $\gamma=\sqrt{\tau_{A}}%
\alpha-\sqrt{\tau_{B}}\beta^{\ast}+\hat{\delta}_{\text{noise}}$,
where the empirical values of $\tau_{A}$ and $\tau_{B}$ are
accessible to the parties from the first-order moments of the
statistics $p(\alpha,\beta,\gamma)$. We assume that Alice is the
encoder and Bob is the decoder, which means that $\alpha$ is
inferred by processing $\beta$ into an optimal estimator. For
large modulation $\varphi\gg1$, Alice and Bob's mutual information
is given by $I_{AB}=\log_{2}(\varphi/\chi)$, where the equivalent
noise $\chi=\chi
(\tau_{A},\tau_{B},\omega_{A},\omega_{B},g,g^{\prime})$ can be
computed from the second-order moments of $p(\alpha,\beta,\gamma)$
(see Supplementary Information for more details).

From the analysis of the second-order statistical moments, Alice
and Bob can derive the secret-key rate of the protocol, which
becomes a simple function of
$\tau_{A}$, $\tau_{B}$ and $\chi$ in the limit of large modulation%
\begin{align}
R(\tau_{A},\tau_{B},\chi)  &  =h\left(
\tfrac{\tau_{A}\chi}{\tau_{A}+\tau
_{B}}-1\right)  -h\left[  \tfrac{\tau_{A}\tau_{B}\chi-(\tau_{A}+\tau_{B})^{2}%
}{|\tau_{A}-\tau_{B}|(\tau_{A}+\tau_{B})}\right] \nonumber\\
&  +\log_{2}\left[
\tfrac{2(\tau_{A}+\tau_{B})}{e|\tau_{A}-\tau_{B}|\chi
}\right]  , \label{RateASYMM}%
\end{align}
where $h(x):=\frac{x+1}{2}\log_{2}\frac{x+1}{2}-\frac{x-1}{2}\log_{2}%
\frac{x-1}{2}$. The asymptotic rate is continuous in
$\tau_{A}=\tau_{B}$,
where it becomes%
\begin{equation}
R(\chi)=h\left(  \tfrac{\chi}{2}-1\right)  +\log_{2}\left[  \tfrac{16}%
{e^{2}\chi(\chi-4)}\right]  . \label{RateSYMM}%
\end{equation}
As typical in QKD, the equivalent noise can be decomposed as $\chi
=\chi_{\text{loss}}+\varepsilon$, where $\chi_{\text{loss}}=2(\tau_{A}%
+\tau_{B})/\tau_{A}\tau_{B}$ is the noise due to loss, while
$\varepsilon$ is the `excess noise'. Thus, the key rate can be
also expressed as $R=R(\tau _{A},\tau_{B},\varepsilon)$.

To study the maximum theoretical performance of the protocol we
set $\varepsilon=0$, therefore restricting Eve to a pure-loss
attack of the links with rate
$R_{\text{loss}}(\tau_{A},\tau_{B}):=R(\tau_{A},\tau_{B},0)$. In
the
symmetric configuration $\tau_{A}=\tau_{B}:=\tau$, we find $R_{\text{loss}%
}(\tau,\tau)\simeq0$ at $\tau\simeq0.84$, so that Alice's and
Bob's distances from a perfectly-in-the-middle relay are limited
to $\simeq$3.8km in standard telecom fibres (0.2dB/km). For this
reason, we consider asymmetric configurations where one of the
links has small loss.

First suppose that Bob's link has small loss ($\tau_{B}\simeq1$).
We find $R_{\text{loss}}\simeq\log_{2}[\tau_{A}/(1-\tau_{A})e]$
which is zero at $\tau_{A}\simeq0.73$. In telecom fibres this is
still restrictive, since Alice's distance from the relay cannot
exceed $6.8$km. By contrast, suppose that Alice's link has small
loss ($\tau_{A}\simeq1$). Now we have $R_{\text{loss}}\simeq
h[(2-\tau_{B})/\tau_{B}]+\log_{2}[\tau_{B}/(1-\tau _{B})e]$, which
goes to zero only for $\tau_{B}\rightarrow0$, corresponding to Bob
arbitrarily far from the relay. Thus, we find that extremely long
distances can be achieved if the relay is sufficiently close to
Alice~\cite{decoder}. These distances are fully quantified in the
next section.

\begin{figure}[ptbh] \vspace{-3.3cm}
\par
\begin{center}
\includegraphics[width=0.6\textwidth] {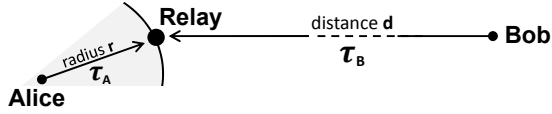}
\end{center}
\par
\vspace{-3.3cm}\caption{Key-distribution via a proximal untrusted
relay, at
short radial distance $r$ from Alice, and distant $d$ from Bob.}%
\label{radius}%
\end{figure}

\textbf{Long distance distribution via proximal
relays}.--~Consider the asymmetric scenario in Fig.~\ref{radius},
where Alice and Bob exploit an untrusted relay at some short
radial distance $r$ from Alice and distance $d$ from Bob. For
pure-loss links ($\varepsilon=0$) and standard fibres
($0.2$dB/km), we can express the rate as $R_{\text{loss}}=R_{\text{loss}%
}(r,d)$. Solving $R_{\text{loss}}=0$ we derive the security
threshold in terms of Bob's maximum distance $d$ for a given value
of $r$.

\begin{figure}[ptbh] \vspace{+0.1cm}
\par
\begin{center}
\includegraphics[width=0.48\textwidth] {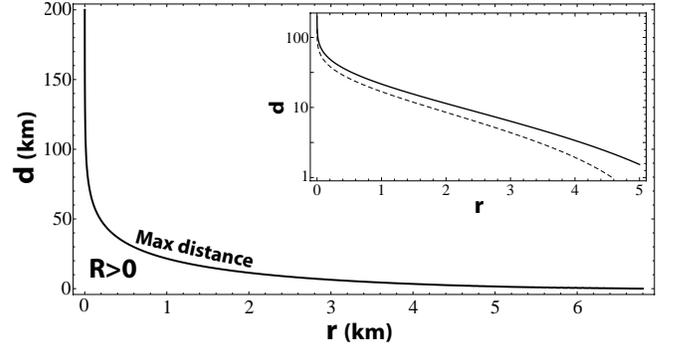}
\end{center}
\par
\vspace{-0.5cm}\caption{Given a relay at distance $r$ from Alice,
the solid line represents the maximum distance $d$ of Bob from
this relay, within which key distribution is possible ($R>0$). The
inset shows Bob's maximum distance (in logarithmic scale) when Eve
performs a coherent Gaussian attack with excess noise
$\varepsilon=0.1$ (dashed line). The corresponding security
threshold is compared with that of the pure-loss attack (solid line).}%
\label{Distances}%
\end{figure}

As we see from Fig.~\ref{Distances}, for a relay sufficiently
close to Alice, Bob can be very far with key distribution being
possible at distances well beyond $100$km. This performance is
very robust to the presence of excess noise $\varepsilon\neq0$,
coming from a coherent Gaussian attack of the links. In this
general case, we can write the rate as $R=R(r,d,\varepsilon)$.
Then, solving $R=0$ for high excess noise
$\varepsilon=0.1$~\cite{Excess} we derive the threshold in the
inset of Fig.~\ref{Distances}. As a result, key distribution is
possible at very long distances even in the presence of very
strong eavesdropping.

\begin{figure}[ptbh] \vspace{-0.1cm}
\par
\begin{center}
\includegraphics[width=0.48\textwidth] {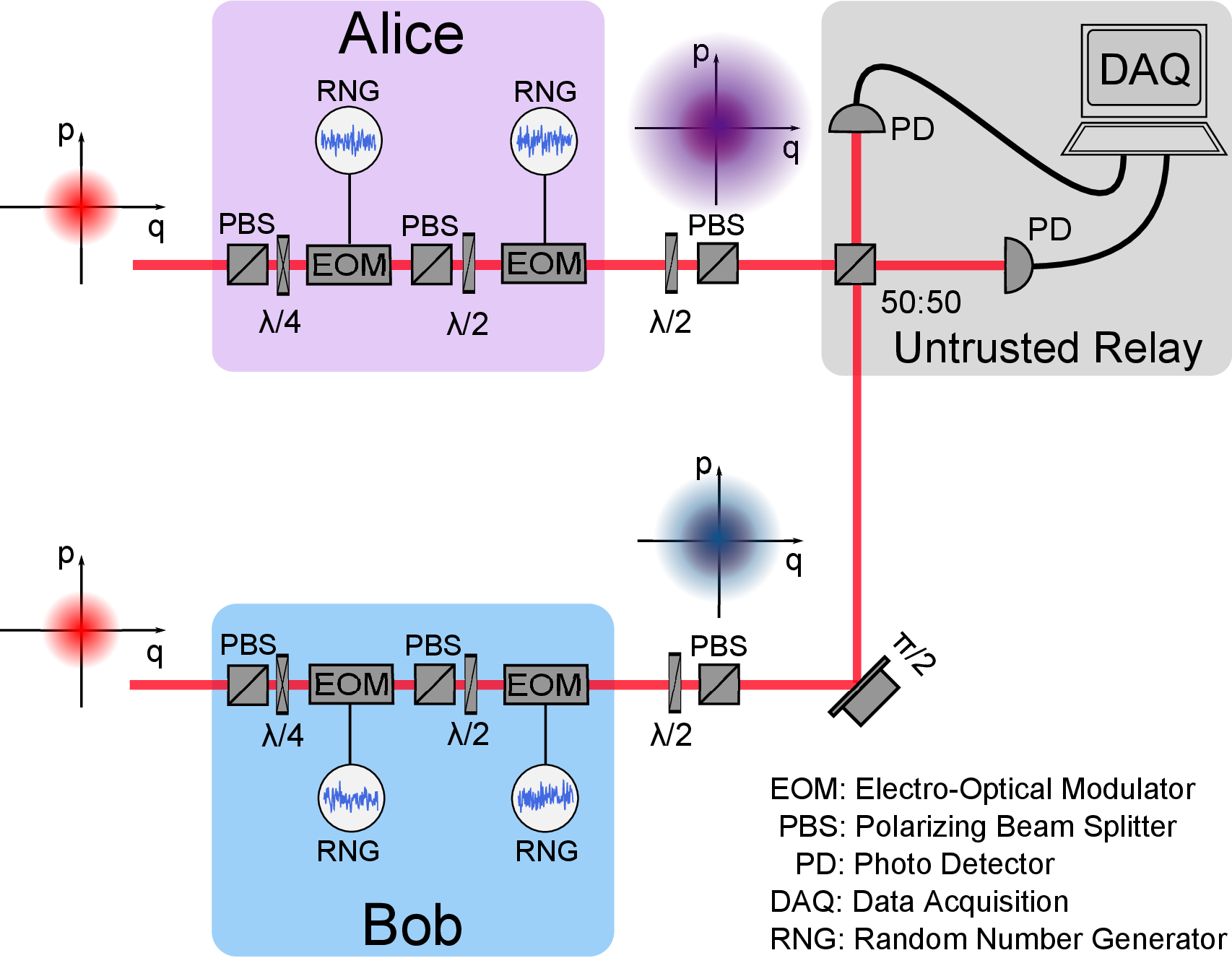}
\end{center}
\par
\vspace{-0.3cm}\caption{Experimental setup. Alice and Bob apply
amplitude and phase modulators to a pair of identical classical
phase-locked bright coherent beams (providing a common local
oscillator). At the output, the two modes emerge
randomly-displaced in the phase space according to a Gaussian
distribution. In particular, Bob's modulation is suitably
attenuated to simulate loss in his link. At the relay, modes are
mixed in a balanced beam splitter and the output ports
photo-detected. Photocurrents are then processed to realize a
continuous-variable Bell measurement (see Supplementary
Information for more details).}%
\label{figSETUP}%
\end{figure}

\textbf{Experimental proof-of-principle}.--~Our theory has been
experimentally confirmed. We have reproduced the asymmetric
configuration of Fig.~\ref{radius}, with $\tau_{A}\simeq1$ and
variable $\tau_{B}$, down to $4\times10^{-4}$ corresponding to
about $170$km in standard optical fibre. A schematic of our
experimental setup is depicted in Fig.~\ref{figSETUP}
.

\begin{figure}[ptbh] \vspace{-1.0cm}
\par
\begin{center}
\includegraphics[width=0.53\textwidth] {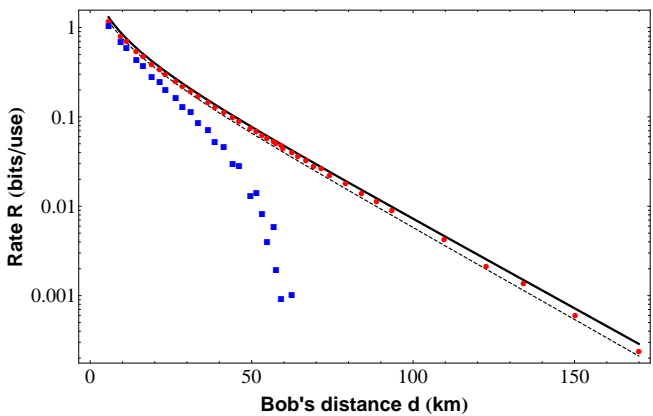}
\end{center}
\par
\vspace{-0.7cm}\caption{Secret-key rate $R$ versus Bob's distance
$d$ from the relay. Experimental points refer to ideal
reconciliation ($\xi=1$, red circles) and realistic reconciliation
($\xi\simeq0.97$, blue squares). For comparison, we also plot the
theoretical rates for a pure-loss attack (solid line) and a
coherent Gaussian attack with excess noise $\varepsilon=0.02$
(dashed line).}%
\label{picEXP}%
\end{figure}

For every experimental point, we have evaluated the second-order
moments of $p(\alpha,\beta,\gamma)$ and computed the experimental
key rate $R=\xi I_{AB}-I_{E}$, with $\xi\leq1$ being the
reconciliation efficiency (current achievable value
$\xi\simeq0.97$~\cite{Jouguet}). Experimental results are plotted
in Fig.~\ref{picEXP}\ and compared with the theoretical
predictions, with excellent agreement. Assuming ideal
reconciliation, the experimental rate approaches the theoretical
limit of the pure-loss attack. Due to imperfections, we have extra
noise in our data which affects the rate approximately in the same
way as a coherent Gaussian attack with excess noise
$\varepsilon\lesssim0.02$. Note that we can potentially reach
$R\simeq10^{-4}$ secret bits per relay use over a link with $34$dB
loss, equivalent to $170$km in standard optical fibre.

Such long distance results are only potential since the current
reconciliation procedures for continuous-variable protocols do not
have unit efficiency (indeed this is the main factor limiting the
distance of continuous variable QKD). By taking this realistic
limitation into account ($\xi\simeq0.97$), we can still reach
remarkably high rates over distances well beyond the typical
connection lengths of a network. As we can see from
Fig.~\ref{picEXP}, we can achieve $R\simeq10^{-2}$ secret bits per
relay use over a link with $10$dB loss, equivalent to $50$km in
fibre.

This result is at least three orders of magnitude higher than that
achievable with discrete-variables over comparable
distances~\cite{Lo,EXP1,EXP2,EXP3}. Implementing our protocol with
a $75$MHz clock rate~\cite{EXP3} would provide a secret-key rate
in the range $75\div750$kbit/s at $50\div60$km, which is
remarkably higher than the value $\lesssim100$bit/s reported in
\cite{EXP3} over similar distances. At the same time, we achieve
the best performances of continuous variable protocols despite the
fact we are removing their point-to-point quantum communication
channel. For instance, working with a $1$MHz clock rate at
$50\div60$km, we can reach a rate of about $1\div 10$kbit/s as in
Ref.~\cite{Grangier2}.

\section{Discussion}

In this work, we have extended the field of continuous-variable
quantum cryptography to a network formulation where two end-users
do not access a direct quantum communication channel but are
forced to connect to an untrusted relay via insecure quantum
links. Despite the possibility that the relay could be fully
corrupted and its links subject to coherent attacks, the end-users
can still extract a secret key. This surprising result comes from
a demanding cryptoanalysis of our model which represents the first
continuous-variable protocol whose rate has been explicitly
computed against a two-mode coherent attack.

An important feature is the simplicity of the relay, which does
not possess any quantum source but just performs a standard
optical measurement, with all the heavy procedures of data
post-processing left to the end-users, fulfilling the idea behind
the end-to-end principle. In particular, the relay implements a
continuous-variable Bell detection which involves highly efficient
photodetectors plus linear optics, whereas the discrete-variable
version of this measurement needs nonlinear elements to operate
deterministically. This feature combined with the use of coherent
states makes the scheme very attractive, guaranteeing both cheap
implementation and high rates.

We have found that the optimal network configuration is asymmetric
with the untrusted relay acting as a proxy server near to one of
the parties. In this case we have experimentally proven that
remarkable rates can be reached, several orders of magnitude
higher than those achievable with qubit-based protocols over
comparable distances. Further improvements in the classical
reconciliation techniques would make the performance of our
protocol even better. From this point of view, our protocol can
already be used for setting up very efficient star networks based
on public access points. Further investigations could involve the
analysis of mixed technology environments where some of the
connections are established at low frequencies (infrared or
microwave) with thermal effects becoming
important~\cite{Weed2010,Weed2012,Weed2013}.

By introducing the mediation of an untrusted and cheap relay, our
work paves the way for extending quantum cryptography to more
advanced models of networks where information is routed end-to-end
through untrusted nodes, whereas current models are strongly based
on computationally demanding point-to-point sessions of key
distribution involving chains of trusted
nodes~\cite{SECOQC,SECOQC2,Tokyo1,Tokyo2}. Indeed our protocol can
already be used to remove trust and reliability from half of the
nodes of a large network, since any chain of nodes between two
end-users can be decomposed into $n+1$ trusted nodes and $n$
untrusted relays (so that only $n$ temporary keys must be
distributed along the chain, instead of $2n$ point-to-point keys).
It is clear that further work is needed to extend the model and
realize a fully untrusted network where only the end-users are
trusted.

\section{Methods}

Theoretical methods, experimental details and data analysis can be
found in the Supplementary Information.

\section*{Acknowledgments}

This work has been supported by EPSRC (grant numbers EP/J00796X/1
and EP/L011298/1), NSERC and the Leverhulme Trust.

%%%%%%%%%%%%%%%%%%%%%%%%%%%%%%%%%%%%%%%%%%%%%%%%%%%%%%%%%%%%%%%%%%%%%%%%%%%%%%%%%%%%%%%%%%%%%%%%%

%\newcounter{S}
\setcounter{section}{0} \setcounter{subsection}{0}
\renewcommand{\bibnumfmt}[1]{[S#1]} \renewcommand{\citenumfont}[1]{S#1}

%\numberwithin{equation}{section}

\begin{center}
{\huge Supplementary Information}

\bigskip
\end{center}

\textbf{Contents of the document}.~In Sec.~\ref{Methods_APP} we
give full details of our theoretical methods
and derivations. We follow the notation of Ref.~\cite{RMPapp}, where $[\hat{q%
},\hat{p}]=2i$ (i.e., $\hslash=2$) so that the vacuum noise is set
to $1$ and $\hat{a}=(\hat{q}+i\hat{p})/2$. In Sec.~\ref{EXP_app}
we thoroughly discuss the experimental setup and the
post-processing of the data, including an analysis of the various
finite-size effects.

\section{Theoretical methods\label{Methods_APP}}

In this section we describe the main theoretical methods used in
our study.
Adopting an entanglement-based representation of the protocol (Sec.~\ref%
{EBased_APP}), we first show how a joint attack of the relay and
the links can be reduced to a Gaussian attack of the links with a
properly-working relay (Sec.~\ref{MID_APP}). Here the key-step is
proving the measurement-device independence for our
continuous-variable protocol, followed by the application of the
known result on the extremality of Gaussian states. In
Sec.~\ref{Comp_APP}, we then describe how to derive the secret-key
rate of the protocol in the presence of an arbitrary Gaussian
attack of the links. In Sec.~\ref{Class_APP} we make our
derivation more specific considering a realistic form of a
coherent Gaussian attack,
deriving an explicit analytical formula for the secret-key rate in Sec.~\ref%
{ANArate_APP}. This rate is minimized and simplified in Sec.~\ref%
{minimal_APP}, and studied for specific configurations in Sec.~\ref{SPEC_APP}%
. Finally, Sec.~\ref{postCMapp} contains the technical derivation
of the post-relay quantum covariance matrix (CM) which is central
for the computation of the rate.

\subsection{Cryptoanalysis of the protocol in the entanglement-based
representation\label{EBased_APP}}

\begin{figure}[ptbh]
\vspace{-1.7cm}
\par
\begin{center}
\includegraphics[width=0.5\textwidth] {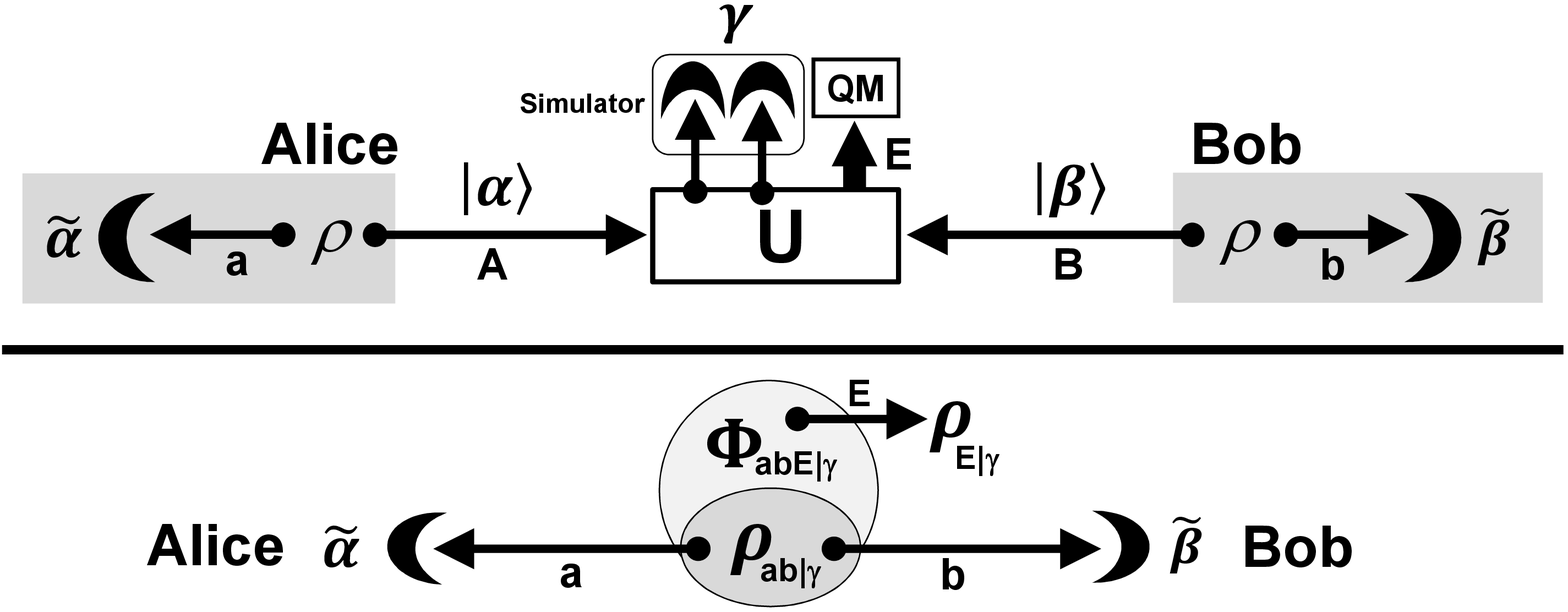}
\end{center}
\par
\vspace{-1.9cm} \caption{(Top).~Entanglement-based representation
of the protocol, where Alice's and Bob's random coherent states
are generated by local heterodyne
detections on EPR states. Outcome variables $\tilde{\protect\alpha}$ and $%
\tilde{\protect\beta}$ are informationally-equivalent to the amplitudes $%
\protect\alpha$ and $\protect\beta$ of the coherent states.
(Bottom)~Conditional scenario after Eve's detection with outcome $\protect%
\gamma$.} \label{puri}
\end{figure}

To study the security of the protocol, we adopt an
entanglement-based representation where each source of coherent
states is realized by an Einstein-Podolsky-Rosen (EPR) state
subject to heterodyne detection. As in
the top panel of Fig.~\ref{puri}, at Alice's station we take an EPR\ state $%
\rho_{aA}$ having zero mean and CM equal to\
\begin{equation}
\mathbf{V}=\left(
\begin{array}{cc}
\mu\mathbf{I} & \sqrt{\mu^{2}-1}\mathbf{Z} \\
\sqrt{\mu^{2}-1}\mathbf{Z} & \mu\mathbf{I}%
\end{array}
\right) ,~\mathbf{Z}:=\left(
\begin{array}{cc}
1 &  \\
& -1%
\end{array}
\right) .   \label{EPRinput}
\end{equation}

By heterodyning mode $a$, Alice remotely prepares a coherent state $%
\left\vert \alpha\right\rangle $ on mode $A$, whose amplitude is
modulated by a complex Gaussian with variance $\varphi=\mu-1$. The
outcome of the measurement $\tilde{\alpha}$ is related to the
projected amplitude $\alpha$ by the relation
\begin{equation}
\tilde{\alpha}=\eta\alpha^{\ast},~\eta:=(\mu+1)(\mu^{2}-1)^{-1/2}.
\label{alphaTILD}
\end{equation}
For large modulation $\mu\gg1$ we have
$\tilde{\alpha}\simeq\alpha^{\ast}$. It is important to note that
the two variables $\alpha$ and $\tilde{\alpha}$ are equivalent
from a information-theoretical point of view, in the sense that
they share the same mutual information with any third variable.

On the other side, Bob's coherent state $\left\vert
\beta\right\rangle $ can be prepared using another EPR state
$\rho_{bB}$ whose mode $b$ is
heterodyned with outcome $\tilde{\beta}=\eta\beta^{\ast}$ (with the limit $%
\tilde{\beta }\simeq\beta^{\ast}$ for $\mu\gg1$). Again, we have
that the outcome variable $\tilde{\beta}$ is informationally
equivalent to $\beta$.

In the top panel of Fig.~\ref{puri} we see that, before the
unitary $U$ and the measurements, the global input state of Alice,
Bob, and Eve is pure and Gaussian (Eve's ancillas are prepared in
vacua). After $U$ and before the measurements, their global output
state is still pure, despite the fact that it could be
non-Gaussian. Since local measurements commute, we can postpone
Alice's and Bob's heterodyne detections after Eve's detection,
whose outcome $\gamma$ is obtained with probability $p(\gamma)$.
Thus, we have the conditional scenario depicted in the bottom
panel of Fig.~\ref{puri}, where Alice, Bob, and Eve share a
conditional state $\Phi_{abE|\gamma}$ with reduced states
$\rho_{ab|\gamma}$ (for Alice and Bob) and $\rho_{E|\gamma}$ (for
Eve). Since Eve performs homodyne detections, $\Phi_{abE|\gamma}$
is pure, so that $\rho_{ab|\gamma}$ and $\rho_{E|\gamma}$ have the
same entropy
\begin{equation}
S(\rho_{ab|\gamma})=S(\rho_{E|\gamma}).
\end{equation}

In the conditional post-relay scheme of Fig.~\ref{puri} (bottom
panel),
Alice encodes information by heterodyning her mode $a$ with result $\tilde{%
\alpha}$. Since heterodyne is a rank-one measurement, it projects $%
\Phi_{abE|\gamma}$ into a pure state
$\Phi_{bE|\gamma\tilde{\alpha}}$ for
Bob and Eve, so that the reduced states $\rho_{b|\gamma\tilde{\alpha}}$ and $%
\rho_{E|\gamma \tilde{\alpha}}$ have the same entropy
\begin{equation}
S(\rho_{b|\gamma\tilde{\alpha}})=S(\rho_{E|\gamma\tilde{\alpha}}).
\label{condENTpuri}
\end{equation}
As a result we have that, conditioned on $\gamma$, Eve's stolen
information
on Alice's variable $\tilde{\alpha}$ is upperbounded by the Holevo quantity%
\begin{equation}
I_{E|\gamma}:=S(\rho_{E|\gamma})-S(\rho_{E|\gamma\tilde{\alpha}%
})=S(\rho_{ab|\gamma})-S(\rho_{b|\gamma\tilde{\alpha}}),
\label{Holquantity}
\end{equation}
which is fully determined by the state $\rho_{ab|\gamma}$.

To retrieve Alice's encoding $\tilde{\alpha}$, Bob heterodynes his
mode $b$
whose output $\tilde{\beta}$ is optimally postprocessed. Conditioned on $%
\gamma$, Alice and Bob's mutual information is given by $I_{AB|\gamma }:=I(%
\tilde{\alpha},\tilde{\beta}|\gamma)=I(\alpha,\beta|\gamma)$,
which is fully determined by $\rho_{ab|\gamma}$. As a result, the
secret-key rate of the protocol is given by the average
\begin{equation}
R=\int d^{2}\gamma~p(\gamma)R_{|\gamma},~R_{|\gamma}:=I_{AB|\gamma
}-I_{E|\gamma}.
\end{equation}
This quantity only depends on the conditional state
$\rho_{ab|\gamma}$ and the statistics of Eve's outcomes
$p(\gamma)$.

\subsection{Measurement-device independence and Gaussian optimality\label%
{MID_APP}}

Starting from the observed distribution
$p(\alpha,\beta,\gamma)=p(\alpha ,\beta|\gamma)p(\gamma)$, Alice
and Bob reconstruct a joint attack as in the top panel of
Fig.~\ref{puri}, which is optimal and compatible with their
statistics. In particular, they compute the post-relay conditional state $%
\rho_{ab|\gamma}$ directly from the conditional probability
$p(\alpha ,\beta|\gamma)$.

Given $p(\alpha,\beta,\gamma)$, and therefore $p(\gamma)$ and
$\rho _{ab|\gamma}$, the rate $R$ is completely determined. This
means that $R$ remains exactly the same if we arbitrarily change
$U$ while preserving the observed statistics
$p(\alpha,\beta,\gamma)$. In particular, we can change $U $ in
such a way that the simulator represents a properly-working relay.
This
can always be done by adding the identity $I=U_{\text{Bell}}U_{\text{Bell}%
}^{\dagger}$, where $U_{\text{Bell}}^{\dagger}$ is absorbed by $U$, while $%
U_{\text{Bell}}$ converts the homodynes of the simulator into a
Bell detection. Thus, compatibly with the observed data, an
arbitrary joint attack $U$\ of the links and the relay can always
be reduced to an attack of the links only, associated with a
properly-working relay.

The next step is exploiting the optimality of Gaussian attacks for
Gaussian protocols~\cite{Raulapp,RMPapp}. For any outcome
$\gamma$, the conditional rate $R_{|\gamma}$ is minimized if we
replace the pure state $\Phi _{abE|\gamma}$\ with a pure Gaussian
state $\Phi_{abE|\gamma}^{G}$ having the same mean value and CM.
Note that, in the top panel of Fig.~\ref{puri}, this is equivalent
to considering $U$ to be a Gaussian unitary, therefore generating
joint Gaussian statistics for the variables $\alpha$, $\beta$ and
$\gamma$. Thus, Alice and Bob can always replace
$p(\alpha,\beta,\gamma)$ with a Gaussian distribution
$p_{G}(\alpha,\beta,\gamma)$ having the same first- and
second-order statistical moments. Adopting this new distribution,
they upperbound Eve's performance replacing her attack with a
Gaussian attack against the links.

\subsection{General computation of the secret-key rate\label{Comp_APP}}

According to the previous analysis, Alice and Bob consider the
worst-case scenario where the two links are subject to a Gaussian
attack, while the relay is properly operated by Eve. In this case,
for any outcome $\gamma$,
the conditional Gaussian state $\rho_{ab|\gamma}$ has always the same CM $%
\mathbf{V}_{ab|\gamma}$ while its mean value varies with $\gamma$.
As typical in Gaussian entanglement swapping~\cite{GaussSWAP},
this random shift in the first moments can be automatically
compensated for by Bob's postprocessing.

As a result we have that $I_{AB|\gamma}$ depends only on the invariant CM $%
\mathbf{V}_{ab|\gamma}$, so that we can write
\begin{equation}
I_{AB}=I_{AB|\gamma}
\end{equation}
for any $\gamma$. Similarly, the entropies in
Eq.~(\ref{Holquantity}) depends only on the invariant CM
$\mathbf{V}_{ab|\gamma}$, so that
\begin{equation}
I_{E}=I_{E|\gamma}
\end{equation}
for any $\gamma$. As a result, the conditional rate $R_{|\gamma}$
is invariant and coincides with the actual (average) rate of the
protocol, i.e., $R=R_{|\gamma}$ for any $\gamma$. Equivalently, we
may write
\begin{equation}
R=I_{AB}-I_{E}.
\end{equation}
Thus, we can theoretically compute the rate $R$ from the quantum CM $\mathbf{%
V}_{ab|\gamma}$ of the post-relay state $\rho_{ab|\gamma}$.

Now it is important to note that this CM can always be determined
by the parties during the data comparison at the end of the
protocol. In particular, it can be derived from the second-order
statistical moments of
the observed data. In fact, from the empirical distribution $%
p(\alpha,\beta,\gamma)$ or its Gaussian version $p_{G}(\alpha,\beta,\gamma)$%
, Alice and Bob derive the relation matrix $\boldsymbol{\Gamma}%
(\alpha,\beta,\gamma)$ and the complex CM $\mathbf{V}(\alpha,\beta,\gamma)$~%
\cite{Pincibono}. Setting $\alpha =(q_{A}+ip_{A})/2$ and $%
\beta=(q_{B}+ip_{B})/2$, these matrices are equivalent to a real
CM for the quadratures
\begin{equation}
\mathbf{V}(q_{A},p_{A},q_{B},p_{B},q_{-},p_{+})=\left(
\begin{array}{cc}
\mathbf{V}_{A\oplus B} & \mathbf{C} \\
\mathbf{C}^{T} & \mathbf{R}%
\end{array}
\right) ,   \label{decogCM}
\end{equation}
where $\mathbf{V}_{A\oplus B}:=\mathbf{V}(q_{A},p_{A})\oplus\mathbf{V}%
(q_{B},p_{B})$ is Alice and Bob's reduced CM, $\mathbf{R}$ is the
CM of the outcomes at the relay, and $\mathbf{C}$ accounts for the
correlations. Given the outcome $\gamma$, the conditional CM of
Alice and Bob can be computed by
Gaussian elimination and reads%
\begin{equation}
\mathbf{V}(q_{A},p_{A},q_{B},p_{B}|\gamma)=\mathbf{V}_{A\oplus B}-\mathbf{CR}%
^{-1}\mathbf{C}^{T}.   \label{GaussELI}
\end{equation}

In the entanglement-based representation of the protocol, we can
easily put the previous CM in terms of Alice's and Bob's
measurement outcomes, respectively,
\begin{equation}
\tilde{\alpha}=(\tilde{q}_{A}+i\tilde{p}_{A})/2,~~\tilde{\beta}=(\tilde{q}%
_{B}+i\tilde{p}_{B})/2.
\end{equation}
In fact, we can write
\begin{equation}
\mathbf{V}(q_{A},p_{A},q_{B},p_{B}|\gamma)=\eta^{-2}\mathbf{V}(\tilde{q}_{A},%
\tilde{p}_{A},\tilde{q}_{B},\tilde{p}_{B}|\gamma),
\label{V1appEX}
\end{equation}
where parameter $\eta$ is defined in Eq.~(\ref{alphaTILD}) and
accounts for the fact that the modulation is finite. Finally, the
connection with the quantum CM is given by
\begin{equation}
\mathbf{V}(\tilde{q}_{A},\tilde{p}_{A},\tilde{q}_{B},\tilde{p}_{B}|\gamma)=%
\mathbf{V}_{ab|\gamma}+\mathbf{I},   \label{classCMoutcomes}
\end{equation}
where $\mathbf{I}$ is the vacuum shot-noise introduced by the
heterodyne detections. The latter relation can be derived
modelling each heterodyne detector as a quantum-limited amplifier
(with gain $2$) followed by a balanced beam splitter (with vacuum
environment) which is conjugately homodyned at the output ports
(one in the position and the other in the momentum quadrature).

Note that for the specific computation of the conditional entropy of Eq.~(%
\ref{condENTpuri}), which enters in Eve's Holevo information and
therefore
the rate, one has to derive the conditional quantum CM\ $\mathbf{V}_{b|\gamma%
\tilde{\alpha}}$ of Bob's state $\rho_{b|\gamma\tilde{\alpha}}$
after Alice's heterodyne detection with outcome $\tilde{\alpha}$.
For this
calculation, we write $\mathbf{V}_{ab|\gamma}$ in the block form%
\begin{equation}
\mathbf{V}_{ab|\gamma}=\left(
\begin{array}{cc}
\mathbf{a} & \mathbf{c} \\
\mathbf{c}^{T} & \mathbf{b}%
\end{array}
\right) ,
\end{equation}
and we apply the formula~\cite{RMPapp}%
\begin{equation}
\mathbf{V}_{b|\gamma\tilde{\alpha}}=\mathbf{b}-\mathbf{c}^{T}(\boldsymbol{%
\mathbf{a}}+\mathbf{I})^{-1}\mathbf{c,}   \label{ff1}
\end{equation}
or equivalently~\cite{BellFORMULAapp}
\begin{equation}
\mathbf{V}_{b|\gamma\tilde{\alpha}}=\mathbf{b}-\zeta^{-1}\mathbf{c}^{T}(%
\boldsymbol{\Omega\mathbf{a}\Omega}^{T}+\mathbf{I})\mathbf{c},
\label{ff2}
\end{equation}
where $\zeta$ is defined in terms of trace and determinant as $\zeta :=\det(%
\mathbf{a)}+\mathrm{Tr}(\mathbf{a)}+1$, and%
\begin{equation}
\boldsymbol{\Omega}=\left(
\begin{array}{cc}
0 & 1 \\
-1 & 0%
\end{array}
\right) .
\end{equation}
Alternatively, we may compute
\begin{equation}
\mathbf{V}_{b|\gamma\tilde{\alpha}}=\mathbf{V}(\tilde{q}_{B},\tilde{p}%
_{B}|\gamma\tilde{\alpha})-\mathbf{I},
\end{equation}
where
$\mathbf{V}(\tilde{q}_{B},\tilde{p}_{B}|\gamma\tilde{\alpha})$
comes
from $\mathbf{V}(\tilde{q}_{A},\tilde{p}_{A},\tilde{q}_{B},\tilde{p}%
_{B}|\gamma)$ after Gaussian elimination of Alice's outcome
variables.

To derive an explicit formula for the secret-key rate, we consider
a realistic form for the coherent Gaussian attack. The
mathematical details of
this attack are fully discussed in the next section. Then, in Sec.~\ref%
{ANArate_APP}\ we will explicitly compute the analytical
expression of the rate.

\subsection{Realistic Gaussian attack against the links\label{Class_APP}}

\begin{figure}[ptbh]
\vspace{-1.1cm}
\par
\begin{center}
\includegraphics[width=0.5\textwidth] {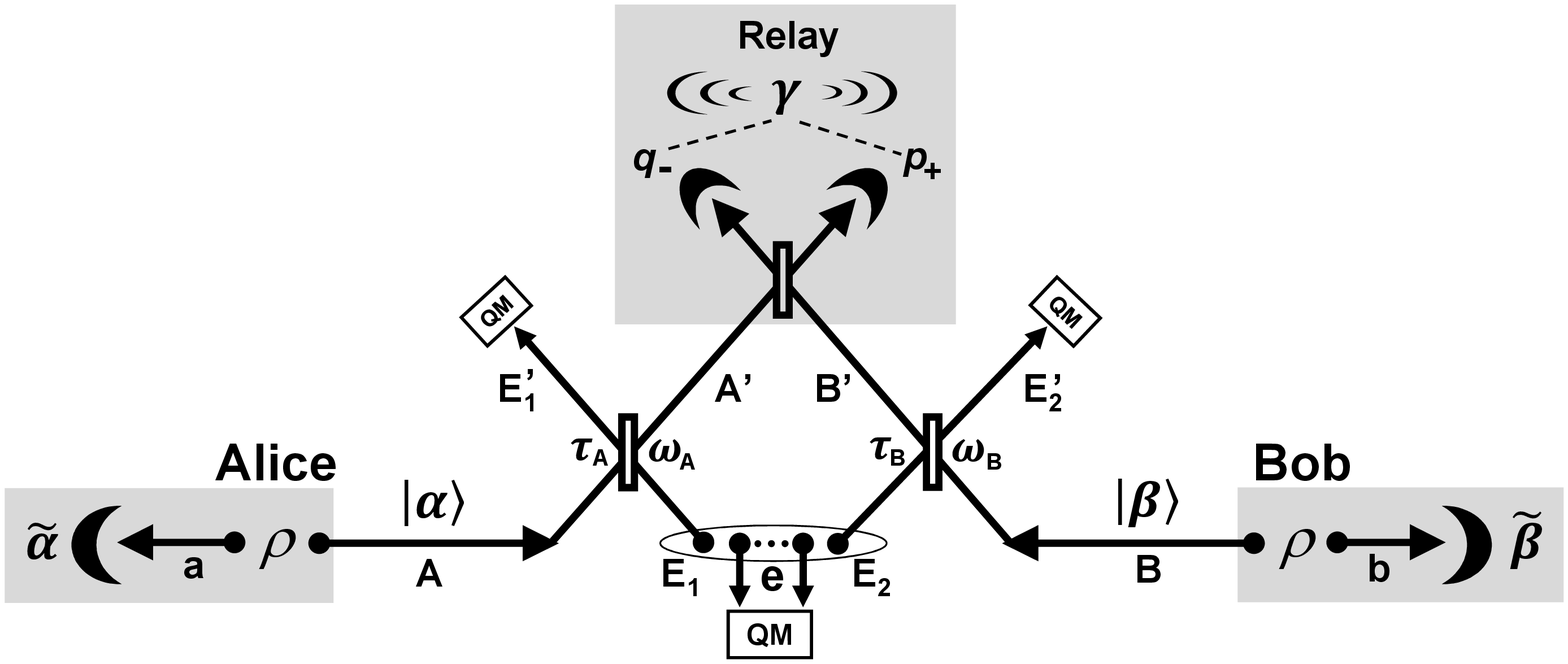}
\end{center}
\par
\vspace{-1.7cm} \caption{Protocol in the entanglement-based
representation. Continuous-variable Bell relay is correctly
operated while a two-mode Gaussian attack is performed against the
two quantum links (see text for more details).} \label{picAPP}
\end{figure}

Consider the scenario depicted in Fig.~\ref{picAPP}, which is the
entanglement-based representation of our protocol subject to a
realistic Gaussian attack against the two links. This two-mode
Gaussian attack is characterized by two beam splitters with
transmissivities $\tau_{A}$ (for Alice's link) and $\tau_{B}$ (for
Bob's link). Using these beam splitters, Eve mixes the incoming
modes, $A$ and $B$, with two ancillary modes, $E_{1}$ and $E_{2}$,
extracted from a reservoir of ancillas. After the beam splitters,
the output ancillas, $E_{1}^{\prime}$ and $E_{2}^{\prime}$, are
stored in a quantum memory (together with all the other ancillas
in the reservoir).

The reduced state $\sigma_{E_{1}E_{2}}$ of $E_{1}$ and $E_{2}$ is
Gaussian
with zero mean-value and CM in the normal form%
\begin{equation}
\mathbf{V}_{E_{1}E_{2}}=\left(
\begin{array}{cc}
\omega_{A}\mathbf{I} & \mathbf{G} \\
\mathbf{G} & \omega_{B}\mathbf{I}%
\end{array}
\right) ,~~%
\begin{array}{c}
\mathbf{I}:=\mathrm{diag}(1,1), \\
\mathbf{G}:=\mathrm{diag}(g,g^{\prime}).%
\end{array}
\label{CMeveAPP}
\end{equation}
Here $\omega_{A}\geq1$ is the variance of the thermal noise introduced by $%
E_{1}$\ in Alice's link, while $\omega_{B}\geq1$\ is the variance
of the thermal noise introduced by $E_{2}$\ in Bob's link. The
correlations between $E_{1}$ and $E_{2}$ are determined by the
block $\mathbf{G}$ whose elements,
$g$ and $g^{\prime}$, must satisfy a set of \textit{bona-fide} conditions~%
\cite{NJP2013app,TwomodePRAapp}. Note that the previous normal
form is very general since any two-mode CM\ can be put in this
form by local Gaussian operations~\cite{RMPapp}.

For given values of thermal noise $\omega _{A},\omega _{B}\geq 1$, Eve's CM $%
\mathbf{V}_{E_{1}E_{2}}$\ is fully determined by the correlation parameters $%
g$ and $g^{\prime }$, which can be represented as a point in a
`correlation plane'. To better describe this plane, we need to
write the bona-fide conditions for $g$ and $g^{\prime }$, which
are derived by imposing the
uncertainty principle~\cite{TwomodePRAapp}%
\begin{equation}
\mathbf{V}_{E_{1}E_{2}}>0,~\nu _{-}^{2}\geq 1,
\end{equation}%
where $\nu _{-}$ is the least symplectic eigenvalue of $\mathbf{V}%
_{E_{1}E_{2}}$. In particular, we have~\cite{RMPapp}
\begin{equation}
2\nu _{-}^{2}=\Delta -\sqrt{\Delta ^{2}-4\det
\mathbf{V}_{E_{1}E_{2}}},
\end{equation}%
where $\Delta =\omega _{A}^{2}+\omega _{B}^{2}+2gg^{\prime }$.

The positivity $\mathbf{V}_{E_{1}E_{2}}>0$ provides the constraints%
\begin{equation}
|g|<\sqrt{\omega_{A}\omega_{B}},~|g^{\prime}|<\sqrt{\omega_{A}\omega_{B}},
\end{equation}
while $\nu_{-}^{2}\geq1$ provides an inequality
$f(\omega_{A},\omega _{B},g,g^{\prime})\geq1$ which is symmetric
with respect to the origin and
the bisector $g^{\prime}=-g$. On the bisector $g^{\prime}=-g$ we have that $%
f\geq1$ corresponds to $|g^{\prime}|\leq\phi$ where%
\begin{equation}
\phi:=\min\left\{
\sqrt{(\omega_{A}-1)(\omega_{B}+1)},\sqrt{(\omega
_{A}+1)(\omega_{B}-1)}\right\} .   \label{gOPT}
\end{equation}

\begin{figure}[tbph]
\vspace{+0.2cm}
\par
\begin{center}
\includegraphics[width=0.30\textwidth] {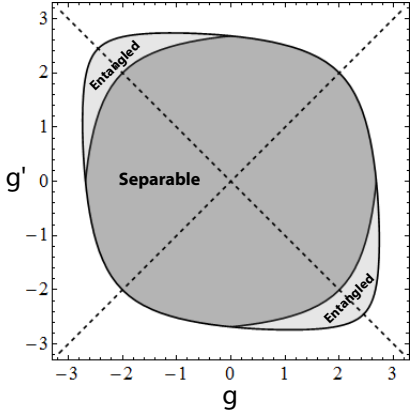}
\end{center}
\par
\vspace{-0.5cm}
\caption{Correlation plane $(g,g^{\prime })$ for $\protect\omega _{A}=5$ and $%
\protect\omega _{B}=2$. Points in the external white area are not
accessible (having correlations too strong to be compatible with
quantum mechanics). Accessible points (i.e., satisfying the
bona-fide conditions) are represented by the delimited region,
which is further divided in sub-regions. The inner darker area
corresponds to separable attacks, while the two peripheral areas
correspond to entangled attacks. The two dashed lines represent
the two bisectors $g^{\prime }=g$ and $g^{\prime }=-g$.}
\label{AttackAPP}
\end{figure}

The previous bona-fide conditions must be satisfied by $g$ and
$g^{\prime }$ for any $\omega _{A},\omega _{B}\geq 1$. They imply
that the correlation plane is accessible only in a limited region
around its origin, whose border
is determined by $\omega _{A}$ and $\omega _{B}$. A numerical example for $%
\omega _{A}=5$ and $\omega _{B}=2$ is provided in
Fig.~\ref{AttackAPP}. The accessible region always forms a
continuous and convex set, apart from the singular case $\omega
_{A}=1$ or $\omega _{B}=1$, where it collapses into its origin
$g^{\prime }=g=0$.

The accessible region of the correlation plane can be further
divided into sub-regions, corresponding to attacks performed with
separable or entangled ancillas $E_{1}$ and $E_{2}$. In fact, we
can easily write the separability condition for Eve's reduced
state $\sigma _{E_{1}E_{2}}$ by considering the least
partially-transposed symplectic eigenvalue $\tilde{\nu}_{-}$ of
its CM $\mathbf{V}_{E_{1}E_{2}}$. This positive eigenvalue is
determined by
\begin{equation}
2\tilde{\nu}_{-}^{2}=\tilde{\Delta}-\sqrt{\tilde{\Delta}^{2}-4\det \mathbf{V}%
_{E_{1}E_{2}}},
\end{equation}%
where $\tilde{\Delta}=\omega _{A}^{2}+\omega _{B}^{2}-2gg^{\prime
}$. Then,
we have $\sigma _{E_{1}E_{2}}$ separable if and only if $\tilde{\nu}%
_{-}^{2}\geq 1$. This corresponds to another inequality
$\tilde{f}(\omega _{A},\omega _{B},g,g^{\prime })\geq 1$ which is
now symmetric with respect to the origin and the bisector
$g^{\prime }=g$. Thus, within the accessible
region, we have an inner area of attacks performed with separable ancillas ($%
\tilde{\nu}_{-}^{2}\geq 1$), that we call `separable attacks', and
two
peripheral areas of attacks performed with entangled ancillas ($\tilde{\nu}%
_{-}^{2}<1$), that we call `entangled attacks'\ (see
Fig.~\ref{AttackAPP}).

Among the separable attacks, the simplest one is the origin of the
plane, where Eve's state is a tensor product $\sigma
_{E_{1}}\otimes \sigma _{E_{2}} $. This attack consists of two
independent entangling cloners, i.e., two beam splitters mixing
the travelling modes with two independent thermal modes, part of
two EPR states. Apart from this singular case, all other separable
attacks have correlated noise, with the two ancillary modes
$E_{1}$ and $E_{2}$ possessing quantum correlations (non-zero
quantum discord) and possibly entangled with extra ancillas
$\mathbf{e}$ in the reservoir.

For the entangled attacks, it is essential to distinguish between
the two
peripheral regions. For attacks in the bottom-right region ($g>0$ and $%
g^{\prime}<0$), Eve injects `good'\ entanglement. She injects EPR
correlations of the type $\hat{q}_{E_{1}}\approx\hat{q}_{E_{2}}$ and $\hat {p%
}_{E_{1}}\approx-\hat{p}_{E_{2}}$, helping the Bell detection
(which
projects on the same kind of correlations $\hat{q}_{A^{\prime}}\approx\hat {q%
}_{B^{\prime}}$ and
$\hat{p}_{A^{\prime}}\approx-\hat{p}_{B^{\prime}}$). We call the
`positive EPR attack'\ the most entangled attack in this region,
which is the bottom-right point with $g=\phi$ and
$g^{\prime}=-\phi$.

By contrast, for attacks in the top-left region ($g<0$ and
$g^{\prime}>0$),
Eve injects `bad'\ entanglement with EPR\ correlations of the type $\hat {q}%
_{E_{1}}\approx-\hat{q}_{E_{2}}$ and $\hat{p}_{E_{1}}\approx\hat{p}_{E_{2}}$%
. These correlations tend to destroy those established by the Bell
detection. In particular, we call the `negative EPR attack'\ the
most entangled state $\sigma_{E_{1}E_{2}}$ in this region, which
is the top-left point with $g=-\phi$ and $g^{\prime}=\phi$.

\subsection{Analytical derivation of the secret-key rate\label{ANArate_APP}}

Having completely characterized the description of the realistic
Gaussian attack against the two links, we can now derive the
corresponding secret-key
rate of the protocol under such an attack. As previously discussed in Sec.~%
\ref{Comp_APP}, this rate is completely determined by the quantum CM\ $%
\mathbf{V}_{ab|\gamma}$ of Alice's and Bob's remote modes $a$ and
$b$ after the action of the relay. The first step is therefore the
computation of the
post-relay CM $\mathbf{V}_{ab|\gamma}$ coming from the scenario in Fig.~\ref%
{picAPP}.

In Sec.~\ref{postCMapp} we explicitly compute%
\begin{align}
\mathbf{V}_{ab|\gamma} & =\left(
\begin{array}{cc}
\mu\mathbf{I} & \mathbf{0} \\
\mathbf{0} & \mu\mathbf{I}%
\end{array}
\right) -(\mu^{2}-1)\times  \notag \\
& \times\left(
\begin{array}{cccc}
\frac{\tau_{A}}{\theta} & 0 & -\frac{\sqrt{\tau_{A}\tau_{B}}}{\theta} & 0 \\
0 & \frac{\tau_{A}}{\theta^{\prime}} & 0 & \frac{\sqrt{\tau_{A}\tau_{B}}}{%
\theta^{\prime}} \\
-\frac{\sqrt{\tau_{A}\tau_{B}}}{\theta} & 0 & \frac{\tau_{B}}{\theta} & 0 \\
0 & \frac{\sqrt{\tau_{A}\tau_{B}}}{\theta^{\prime}} & 0 & \frac{\tau_{B}}{%
\theta^{\prime}}%
\end{array}
\right) ,   \label{VabGamma}
\end{align}
where we have set%
\begin{equation}
\theta:=\left( \tau_{A}+\tau_{B}\right)
\mu+\lambda,~\theta^{\prime }:=\left( \tau_{A}+\tau_{B}\right)
\mu+\lambda^{\prime},   \label{teta2}
\end{equation}
and the lambdas are defined as%
\begin{equation}
\lambda:=\kappa-ug>0,~\lambda^{\prime}:=\kappa+ug^{\prime}>0,
\label{lambda2}
\end{equation}
with
\begin{align}
\kappa & :=(1-\tau_{A})\omega_{A}+(1-\tau_{B})\omega_{B},  \label{kap1} \\
u & :=2\sqrt{(1-\tau_{A})(1-\tau_{B})}.   \label{kap2}
\end{align}
In the limit of high modulation $\mu\gg1$, the symplectic spectrum~\cite%
{RMPapp} of $\mathbf{V}_{ab|\gamma}$ takes the asymptotic
expressions
\begin{equation}
\{\nu_{1},\nu_{2}\}=\left\{ \frac{|\tau_{A}-\tau_{B}|}{\tau_{A}+\tau_{B}}%
\mu,~\frac{\sqrt{\lambda\lambda^{\prime}}}{|\tau_{A}-\tau_{B}|}\right\} ~~%
\text{for }\tau_{A}\neq\tau_{B},   \label{spectrum1}
\end{equation}
and%
\begin{equation}
\{\nu_{1},\nu_{2}\}=\left\{ \sqrt{\frac{\lambda\mu}{2\tau_{B}}},~\sqrt {%
\frac{\lambda^{\prime}\mu}{2\tau_{B}}}\right\} ~~\text{for }%
\tau_{A}=\tau_{B}.   \label{spectrum2}
\end{equation}

For the next calculations it is useful to compute the CMs of Bob's
reduced state $\rho_{b|\gamma}$ and Bob's state
$\rho_{b|\gamma\tilde{\alpha}}$
conditioned on Alice's detection. These CMs can easily be derived from $%
\mathbf{V}_{ab|\gamma}$. In particular, we have%
\begin{align}
\mathbf{V}_{b|\gamma} & =\left(
\begin{array}{cc}
\mu-\frac{(\mu^{2}-1)\tau_{B}}{\theta} & 0 \\
0 & \mu-\frac{(\mu^{2}-1)\tau_{B}}{\theta^{\prime}}%
\end{array}
\right) , \\
\mathbf{V}_{b|\gamma\tilde{\alpha}} & =\left(
\begin{array}{cc}
\mu-\frac{(\mu^{2}-1)\tau_{B}}{\tau_{A}+\tau_{B}\mu+\lambda} & 0 \\
0 & \mu-\frac{(\mu^{2}-1)\tau_{B}}{\tau_{A}+\tau_{B}\mu+\lambda^{\prime}}%
\end{array}
\right) ,   \label{conCMeq}
\end{align}
with$\ \mathbf{V}_{b|\gamma\tilde{\alpha}}$ having the symplectic
eigenvalue
\begin{equation}
\nu=\tau_{B}^{-1}\sqrt{(\tau_{A}+\lambda)(\tau_{A}+\lambda^{\prime})}~~\text{%
for~}\mu\gg1.   \label{niCONDITIONAL}
\end{equation}

\subsubsection{Alice and Bob's mutual information\label{mutual_APP}}

Having determined the post-relay CM, we can now proceed with the
derivation
of the rate. First, let us compute the mutual information of Alice and Bob $%
I_{AB}=I(\tilde{\alpha},\tilde{\beta}|\gamma)$. The outcomes
$\tilde{\alpha}$ and $\tilde{\beta}$\ of Alice's and Bob's
detectors are associated with the classical CM given in
Eq.~(\ref{classCMoutcomes}). In particular, Bob's reduced CM is
equal to
\begin{equation}
\mathbf{V}_{B|\gamma}:=\mathbf{V}(\tilde{q}_{B},\tilde{p}_{B}|\gamma )=%
\mathbf{V}_{b|\gamma}+\mathbf{I}.
\end{equation}
Then, Bob's CM\ conditioned to Alice's outcome $\tilde{\alpha}$ is%
\begin{equation}
\mathbf{V}_{B|\gamma\tilde{\alpha}}:=\mathbf{V}(\tilde{q}_{B},\tilde{p}%
_{B}|\gamma\tilde{\alpha})=\mathbf{V}_{b|\gamma\tilde{\alpha}}+\mathbf{I}.
\end{equation}

The previous CMs are all diagonal
$\mathbf{V}=\mathrm{diag}(V^{q},V^{p})$. Since the two quadratures
are modulated independently, the mutual information $I_{AB}$ is
given by the sum of the two terms, one for each
quadrature. Explicitly, we have%
\begin{equation}
I_{AB}=\frac{1}{2}\log_{2}\frac{V_{B|\gamma}^{q}}{V_{B|\gamma\tilde{\alpha}%
}^{q}}+\frac{1}{2}\log_{2}\frac{V_{B|\gamma}^{p}}{V_{B|\gamma\tilde{\alpha}%
}^{p}}=\frac{1}{2}\log_{2}\Sigma,   \label{mutuaINab}
\end{equation}
where%
\begin{equation}
\Sigma:=\frac{1+\det\mathbf{V}_{b|\gamma}+\mathrm{Tr}\mathbf{V}_{b|\gamma}}{%
1+\det\mathbf{V}_{b|\gamma\tilde{\alpha}}+\mathrm{Tr}\mathbf{V}_{b|\gamma%
\tilde{\alpha}}}.   \label{sigmaAPP}
\end{equation}

We can always re-write the mutual information in terms of a
signal-to-noise
ratio as%
\begin{equation}
I_{AB}=\log_{2}\frac{\mu}{\chi}=\log_{2}\frac{\varphi+1}{\chi},
\label{ABmutual}
\end{equation}
where the equivalent noise $\chi=\mu\Sigma^{-1/2}$ can be computed from $%
\Sigma$, which in turn depends on the post-relay CM $\mathbf{V}_{ab|\gamma}$%
, known to the parties from the observed statistics
$p(a,\beta|\gamma)$. For
large modulation $\mu\gg1$, we derive%
\begin{equation}
\chi=\frac{\tau_{A}+\tau_{B}}{\tau_{A}\tau_{B}}\sqrt{(\tau_{A}+\tau
_{B}+\lambda)(\tau_{A}+\tau_{B}+\lambda^{\prime})},
\label{EqNOISE}
\end{equation}
where the equivalent noise is expressed in terms of all parameters
of the attack, i.e.,
$\chi=\chi(\tau_{A},\tau_{B},\omega_{A},\omega_{B},g,g^{\prime
})$.

\subsubsection{Eve's Holevo information\label{Holevo_APP}}

In order to bound Eve's stolen information on Alice's outcome variable $%
\tilde{\alpha}$, we use the Holevo quantity
\begin{equation}
I_{E}=S(\rho_{ab|\gamma})-S(\rho_{b|\gamma\tilde{\alpha}}).
\label{HolINFOapp}
\end{equation}
The first entropy term $S(\rho_{ab|\gamma})$ can be computed from
the symplectic spectrum $\{\nu_{1},\nu_{2}\}$ of
$\mathbf{V}_{ab|\gamma}$. In particular, we have
\begin{equation}
S(\rho_{ab|\gamma})=h(\nu_{1})+h(\nu_{2}),   \label{totEVE}
\end{equation}
where the $h$-function is defined as%
\begin{align}
h(x) & :=\left( \tfrac{x+1}{2}\right) \log_{2}\left(
\tfrac{x+1}{2}\right) -\left( \tfrac{x-1}{2}\right) \log_{2}\left(
\tfrac{x-1}{2}\right) ,
\label{h_entropic} \\
& \simeq\log_{2}\left( \frac{ex}{2}\right) +O\left( \frac{1}{x}\right) ~%
\text{for }x\gg1.
\end{align}

For large modulation ($\mu\gg1$), the spectrum
$\{\nu_{1},\nu_{2}\}$ is given by Eqs.~(\ref{spectrum1})
and~(\ref{spectrum2}). Correspondingly, we
derive the following asymptotic formula for the entropy%
\begin{equation}
S(\rho_{ab|\gamma})=\left\{
\begin{array}{c}
S_{\neq}~~\text{for~}\tau_{A}\neq\tau_{B}, \\
\\
S_{=}~~\text{for~}\tau_{A}=\tau_{B},%
\end{array}
\right.
\end{equation}
where%
\begin{align}
S_{\neq} & :=h\left( \frac{\sqrt{\lambda\lambda^{\prime}}}{%
|\tau_{A}-\tau_{B}|}\right) +\log_{2}\left[ \frac{e|\tau_{A}-\tau_{B}|\mu}{%
2(\tau _{A}+\tau_{B})}\right] , \\
S_{=} & :=\log_{2}\left( \frac{e^{2}\sqrt{\lambda\lambda^{\prime}}\mu }{%
8\tau_{B}}\right) .
\end{align}
For the second entropy term in Eq.~(\ref{HolINFOapp}), we have
$S(\rho _{b|\gamma\tilde{\alpha}})=h(\nu)$, where the symplectic
eigenvalue $\nu$ takes the asymptotic expression given in
Eq.~(\ref{niCONDITIONAL}).

Here it is important to note that $S(\rho_{ab|\gamma})$ is continuous in $%
\tau_{A}=\tau_{B}$. In other words, we have
\begin{equation}
S_{\neq}\rightarrow S_{=},\text{~~for
}\tau_{A}\rightarrow\tau_{B}\text{.}
\end{equation}
This continuity is inherited by the Holevo information $I_{E}$ in Eq.~(\ref%
{HolINFOapp}) and by the rate computed in the next section.

\subsubsection{Asymptotic secret-key rate\label{Rate_APP}}

Using the previous formulas for Alice and Bob's mutual information
and Eve's
Holevo information, we can derive the secret-key rate of the protocol $%
R=I_{AB}-I_{E}$ for large modulation $\mu\gg1$. This rate is
expressed in
terms of all parameters of the attack $R=R(\tau_{A},\tau_{B},\omega_{A},%
\omega_{B},g,g^{\prime})$. For $\tau_{A}\neq\tau_{B}$, we find%
\begin{equation}
R=h(\nu)-h\left( \frac{\sqrt{\lambda\lambda^{\prime}}}{|\tau_{A}-\tau_{B}|}%
\right) +\log_{2}\left[ \frac{2(\tau_{A}+\tau_{B})}{e|\tau_{A}-\tau _{B}|\chi%
}\right] ,   \label{R1text}
\end{equation}
with continuous limit in $\tau_{A}=\tau_{B}:=\tau$, where it becomes%
\begin{equation}
R=h(\nu)+\log_{2}\left( \frac{8\tau}{e^{2}\chi\sqrt{\lambda\lambda^{\prime}}}%
\right) .
\end{equation}
Here the lambda parameters $\lambda$ and $\lambda^{\prime}$ are
given in
Eq.~(\ref{lambda2}), the eigenvalue $\nu$ is expressed by Eq.~(\ref%
{niCONDITIONAL}), and the equivalent noise\ $\chi$ is given in Eq.~(\ref%
{EqNOISE}).

\subsection{Minimum secret-key rate\label{minimal_APP}}

The general formula in Eq.~(\ref{R1text}) is given in terms of all
parameters of the attack $\tau_{A}$, $\tau_{B}$, $\omega_{A}$,
$\omega_{B}$, $g$, and $g^{\prime}$. While $\tau_{A}$ and
$\tau_{B}$ are easily accessible from the first-order moments, the
remaining parameters could be inaccessible
to the parties, since they generally mix in determining the CM $\mathbf{V}%
_{ab|\gamma}$ of Eq.~(\ref{VabGamma}). For this practical reason,
it is important to express the rate in terms of fewer parameters.

\subsubsection{Minimization of the rate at fixed thermal noise}

Let us start by assuming that Alice and Bob only know the transmissivities ($%
\tau _{A}$ and $\tau _{B}$) and the thermal noise affecting each link ($%
\omega _{A}$ and $\omega _{B}$). Given these four parameters, we
minimize the rate $R(\tau _{A},\tau _{B},\omega _{A},\omega
_{B},g,g^{\prime })$ of Eq.~(\ref{R1text}) over all physical
values of the correlation parameters, i.e., over all accessible
points in the correlation plane $(g,g^{\prime })$. In our
analysis, it is sufficient to assume $\tau _{A}\neq \tau _{B}$,
since we can extend the result to the symmetric configuration
$\tau _{A}=\tau _{B}$ by continuity.

\begin{figure}[ptbh]
\vspace{+0.1cm}
\par
\begin{center}
\includegraphics[width=0.35\textwidth] {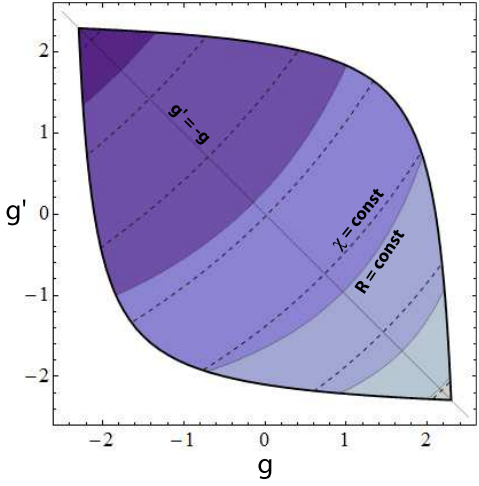}
\end{center}
\par
\vspace{-0.5cm} \caption{\textbf{Minimization at fixed thermal
noise}. The general rate $R$
of Eq.~(\protect\ref{R1text}) is plotted in the correlation plane $%
(g,g^{\prime})$ for $\protect\tau_{A}=0.9$, $\protect\tau_{B}=0.5$, and $%
\protect\omega_{A}=\protect\omega_{B}=2.5$. The non-white area is
the set of accessible points, which is convex and symmetric with
respect to the bisector $g^{\prime}=-g$. Within the accessible
set, darker regions correspond to lower values of the rate. The
rate is symmetric with respect to the bisector $g^{\prime}=-g$,
and decreasing towards the extremal top-left point of the
accessible set (behavior is generic for any choice of the
parameters $\protect\tau_{A},$ $\protect\tau_{B},$
$\protect\omega_{A}$ and$~\protect\omega_{B}$).
\textbf{Mnimization at fixed equivalent noise}. We compare the
iso-noise curves ($\protect\chi$=\textrm{const}) with the more
concave iso-rate curves ($R$=\textrm{const}).} \label{Concavity}
\end{figure}

First it is important to note that the rate $R$ in
Eq.~(\ref{R1text}) depends on the correlation parameters
$(g,g^{\prime})$ only via the lambdas, $\lambda$ and
$\lambda^{\prime}$, specified by Eq.~(\ref{lambda2}). Since $R$ is
invariant under permutation
$\lambda\leftrightarrow\lambda^{\prime}$, it is symmetric with
respect to the bisector $g^{\prime}=-g$. This symmetry is also
shown in the numerical example given in Fig.~\ref{Concavity}.

In the correlation plane, the set of accessible points is a convex
set and symmetric with respect to the bisector $g^{\prime }=-g$.
Combining this topology with the symmetry of the rate $R$ allows
us to restrict its minimization to the accessible points along the
bisector $g^{\prime }=-g$. Setting $g^{\prime }=-g$ corresponds to
setting $\lambda ^{\prime }=\lambda
=\kappa +ug^{\prime }$ in the rate of Eq.~(\ref{R1text}), which gives%
\begin{equation}
R(g^{\prime }=-g)=H(\tau _{A},\tau _{B},\lambda )+L(\tau _{A},\tau
_{B},\lambda ),
\end{equation}%
where
\begin{align}
H(\tau _{A},\tau _{B},\lambda )& :=h\left( \frac{\tau _{A}+\lambda
}{\tau _{B}}\right) -h\left( \frac{\lambda }{|\tau _{A}-\tau
_{B}|}\right) ,
\label{capH} \\
L(\tau _{A},\tau _{B},\lambda )& :=\log _{2}\left[ \frac{2\tau _{A}\tau _{B}%
}{e|\tau _{A}-\tau _{B}|(\tau _{A}+\tau _{B}+\lambda )}\right] .
\label{capL}
\end{align}

It is easy to check that $H$ and $L$ are minimized by maximizing
$\lambda $ which, in turn, corresponds to maximizing $g^{\prime
}$. As we know from Sec.~\ref{Class_APP}, the maximal accessible
value of $g^{\prime }$ along the bisector $g^{\prime }=-g$ is
given by $\phi $ in Eq.~(\ref{gOPT}). Thus,
at fixed transmissivities ($\tau _{A}$ and $\tau _{B}$) and thermal noise ($%
\omega _{A}$ and $\omega _{B}$), the optimal coherent attack is given by $%
g^{\prime }=-g=\phi $, which is the extremal top-left point in the
accessible region of the correlation plane. As already discussed in Sec.~\ref%
{Class_APP}, this entangled attack is a `negative EPR attack'\
where Eve injects EPR correlations in the links which tend to
destroy the effect of
the Bell detection. By contrast, $H$ and $L$ are maximized by minimizing $%
g^{\prime }$, whose minimum accessible value along the bisector
$g^{\prime }=-g$ is given by $-\phi $. Thus, the rate is maximum
in the extremal bottom-right point of the accessible region,
corresponding to the `positive EPR attack', where Eve injects EPR
correlations helping the Bell detection.

Assuming that Eve performs the optimal attack, i.e., the negative
EPR\ attack, the minimum rate of our protocol is equal to
\begin{gather}
R(\tau_{A},\tau_{B},\omega_{A},\omega_{B})=h\left( \frac{\tau_{A}+\lambda_{%
\text{opt}}}{\tau_{B}}\right) -h\left( \frac{\lambda_{\text{opt}}}{%
|\tau_{A}-\tau_{B}|}\right)  \notag \\
+\log_{2}\left[ \frac{2\tau_{A}\tau_{B}}{e|\tau_{A}-\tau_{B}|(\tau_{A}+%
\tau_{B}+\lambda_{\text{opt}})}\right] ,   \label{mainRATE}
\end{gather}
where%
\begin{equation}
\lambda_{\text{opt}}:=\kappa+u\phi,
\end{equation}
with $\kappa$ and $u$ defined in Eqs.~(\ref{kap1})
and~(\ref{kap2}). This
rate is continuous in $\tau_{A}=\tau_{B}:=\tau$, where it becomes%
\begin{equation}
R(\tau,\omega_{A},\omega_{B})=h\left( \frac{\tau+\lambda_{\text{opt}}}{\tau }%
\right) +\log_{2}\left[ \frac{4\tau^{2}}{e^{2}(2\tau+\lambda_{\text{opt}%
})\lambda_{\text{opt}}}\right] .   \label{mainRATEconfsym}
\end{equation}

As evident from Fig.~\ref{Concavity}, the negative EPR\ attack
clearly outperforms the collective entangling-cloner attack
(corresponding to the origin of the plane $g^{\prime}=g=0$). From
this point of view, the security analysis of our protocol is
clearly more complex compared to previous literature on
continuous-variable quantum cryptography.

\subsubsection{Minimization of the rate at fixed equivalent noise}

Despite the complexity of the protocol, it is remarkable that we
can further simplify its rate by expressing it in terms of three
parameters only. In fact, suppose that Alice and Bob only access
the values of the
transmissivities ($\tau_{A}$ and $\tau_{B}$) and the equivalent noise $\chi$%
. This minimal knowledge on the attack is always guaranteed. In
fact, as
discussed in Sec.~\ref{ANArate_APP} and specifically regarding Eq.~(\ref%
{ABmutual}), we have that the equivalent noise is uniquely
determined by the post-relay CM\ $\mathbf{V}_{ab|\gamma}$ which
is, in turn, reconstructed\ from the empirical statistics
$p(\alpha,\beta|\gamma)$. Thus, given the three accessible
parameters $\tau_{A}$, $\tau_{B}$ and $\chi$, we minimize the
general rate $R$ of Eq.~(\ref{R1text}), finding a simple formula
for the minimum rate $R(\tau_{A},\tau_{B},\chi)$.

For this constrained minimization we can construct a Lagrangian
and look for its critical points (procedure is involved and not
reported here). More intuitively, we can find the minimum by
directly comparing the rate and the equivalent noise on the
correlation plane. On this plane, we plot the curves with the same
rate $R$=\textrm{const} (iso-rate curves) and the curves with the
same equivalent noise $\chi$=\textrm{const} (iso-noise curves). As
we can see from the numerical example in Fig.~\ref{Concavity}, the
iso-rate curves are more concave than the iso-noise curves.

Combining this different concavity with the fact that the rate is
decreasing from the bottom-right point to the top-left point of
the accessible region, we find that the rate is minimized at the
intersection of each iso-noise curve with the bisector
$g^{\prime}=-g$. In other words, moving along an iso-noise curve
($\chi$=\textrm{const}) the rate is minimized at the point
where $g^{\prime}=-g$. Therefore we set $g^{\prime}=-g$ in the rate $%
R(\tau_{A},\tau_{B},\omega_{A},\omega_{B},g,g^{\prime})$ and we
express the
result in terms of the three basic parameters $\tau_{A}$, $\tau_{B}$ and $%
\chi$.

Given these parameters the minimum rate is equal to%
\begin{align}
R(\tau_{A},\tau_{B},\chi) & =h\left( \tfrac{\tau_{A}\chi}{\tau_{A}+\tau _{B}}%
-1\right) -h\left[ \tfrac{\tau_{A}\tau_{B}\chi-(\tau_{A}+\tau_{B})^{2}}{%
|\tau_{A}-\tau_{B}|(\tau_{A}+\tau_{B})}\right]  \notag \\
& +\log_{2}\left[ \frac{2(\tau_{A}+\tau_{B})}{e|\tau_{A}-\tau_{B}|\chi }%
\right] ,   \label{Rate_tA_tB_chi}
\end{align}
which is continuous in $\tau_{A}=\tau_{B}$, where it becomes%
\begin{equation}
R(\chi)=h\left( \frac{\chi}{2}-1\right) +\log_{2}\left[ \frac{16}{%
e^{2}\chi(\chi-4)}\right] .
\end{equation}
These latter two formulas represent our main theoretical result
and coincide with Eqs.~(2) and~(3) of the main text.

We can always decompose the equivalent noise as
\begin{equation}
\chi=\chi_{\text{loss}}(\tau_{A},\tau_{B})+\varepsilon,
\end{equation}
where
\begin{equation}
\chi_{\text{loss}}(\tau_{A},\tau_{B}):=\left. \chi\right\vert
_{\omega
_{A}=\omega_{B}=1}=\frac{2(\tau_{A}+\tau_{B})}{\tau_{A}\tau_{B}}
\end{equation}
is that part of the noise due to loss only, with the extra part
$\varepsilon$ known as the `excess noise'. Adopting this
decomposition of the noise, the rate of Eq.~(\ref{Rate_tA_tB_chi})
can be re-written as $R=R(\tau_{A},\tau _{B},\varepsilon)$.

For the specific case $\varepsilon=0$, we have a pure-loss attack
of the links, where Eve's beam splitters mix the incoming modes
with vacuum modes. The study of this simple attack is useful to
estimate the maximum performance of the protocol (which clearly
worsens for $\varepsilon>0$). In
a pure-loss attack, the minimum rate $R(\tau_{A},\tau_{B},0)$ simplifies to%
\begin{align}
R(\tau_{A},\tau_{B}) & =h\left( \frac{2-\tau_{B}}{\tau_{B}}\right)
-h\left(
\frac{2-\tau_{A}-\tau_{B}}{|\tau_{A}-\tau_{B}|}\right)  \notag \\
& +\log_{2}\left(
\frac{\tau_{A}\tau_{B}}{e|\tau_{A}-\tau_{B}|}\right) ,
\label{pLOSSrate}
\end{align}
which is continuous in $\tau_{A}=\tau_{B}:=\tau$, where it becomes
\begin{equation}
R(\tau)=h\left( \frac{2-\tau}{\tau}\right) +\log_{2}\left[ \frac{\tau^{2}}{%
e^{2}(1-\tau)}\right] .   \label{pLOSSrateSYM}
\end{equation}
One can easily check that Eq.~(\ref{pLOSSrate}) can also be
achieved by specifying the previous rates of Eqs.~(\ref{R1text})
and (\ref{mainRATE}) to the specific case of pure-loss
($\omega_{A}=\omega_{B}=1$).

\begin{figure}[ptbh]
\vspace{+0.1cm}
\par
\begin{center}
\includegraphics[width=0.33\textwidth] {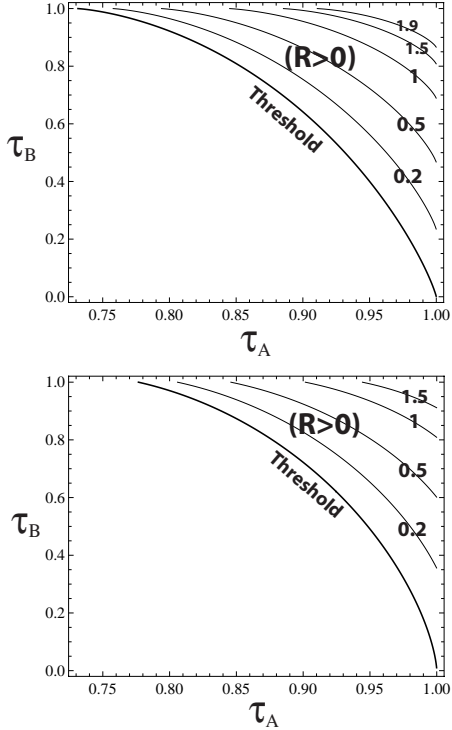}
\end{center}
\par
\vspace{-0.5cm} \caption{Secret-key rate $R$ (bits per relay use)
as a function of the two
transmissivities $\protect\tau_{A}$ and $\protect\tau_{B}$. (Top) Rate $R(%
\protect\tau _{A},\protect\tau_{B})$ for a pure-loss attack ($\protect%
\varepsilon=0$). (Bottom) Rate $R(\protect\tau_{A},\protect\tau_{B},\protect%
\varepsilon)$\ for a coherent Gaussian attack with excess noise $\protect%
\varepsilon=0.1$.} \label{ExcessPlot}
\end{figure}

In the top panel of Fig.~\ref{ExcessPlot}, we have plotted $%
R(\tau_{A},\tau_{B})$ as a function of the two transmissivities.
We see that $R>0$ occurs above a certain threshold which is
asymmetric in the plane. In particular, for $\tau_{A}$ close to
$1$ we see that $\tau_{B}$ can be close to zero, identifying the
optimal configuration of the protocol. Assuming standard optical
fibres ($0.2$dB/km), this corresponds to having Alice close to the
relay while Bob can be very far away as also shown in Fig.~4\ in
the main text. These results are proven to be robust with respect
to the presence of excess noise $\varepsilon>0$. This can be seen
in the bottom
panel of Fig.~\ref{ExcessPlot}, where we plot $R(\tau_{A},\tau_{B},%
\varepsilon)$ for the high numerical value $\varepsilon=0.1$.

\subsection{Minimum rate in limit configurations\label{SPEC_APP}}

According to Fig.~\ref{ExcessPlot}, the most interesting scenario
is when the protocol is implemented in an asymmetric fashion with
small loss in Alice's link ($\tau_{A}\simeq1$). Therefore, let us
analyze what happens in
the limit for $\tau_{A}\rightarrow1$. Assuming a fixed loss rate (e.g., $0.2$%
dB/km), this corresponds to Alice approaching the relay (note that
even if the Alice-relay distance becomes negligible, the relay
must still be considered as a different device which is
potentially operated by Eve). Taking the limit
$\tau_{A}\rightarrow1$ in Eq.~(\ref{Rate_tA_tB_chi}), we
find%
\begin{align}
R_{\tau_{A}\rightarrow1} & =h\left[ \frac{1+(1-\tau_{B})\omega_{B}}{\tau_{B}}%
\right] -h(\omega_{B})  \label{RateRR} \\
& +\log_{2}\left\{
\frac{2\tau_{B}}{e(1-\tau_{B})[1+\tau_{B}+(1-\tau
_{B})\omega_{B}]}\right\} .  \notag
\end{align}
This is a function of $\tau_{B}$ and $\omega_{B}$ only, as
expected since the coherent Gaussian attack collapses into an
entangling cloner attack
affecting Bob's link only (with transmissivity $\tau_{B}$ and thermal noise $%
\omega_{B}$).

It is important to note that the rate of Eq.~(\ref{RateRR}) is
identical to the reverse-reconcilation rate of a point-to-point
`no-switching'\ protocol from Alice to Bob which is based on
random coherent states and heterodyne
detection~\cite{NoSwitchapp} (see the Supplementary Information of Ref.~\cite%
{TwowayPROTOCOLapp} for checking the asymptotic analytical
expression of this rate). It is remarkable that the security
performance of such a powerful protocol can be realized in the
absence of a direct link between Alice and Bob by exploiting the
intermediation of an untrusted relay in the proximity of Alice,
i.e., the encoder of the secret information. Thanks to the
equivalence with reverse reconciliation, we can achieve remarkably
long
distances for Bob. Ideally, if Bob's link were affected by pure loss only ($%
\omega_{B}=1$), then we would have
\begin{equation}
R_{\tau_{A}\rightarrow1}^{\text{loss}}=h\left( \frac{2-\tau_{B}}{\tau_{B}}%
\right) +\log_{2}\left[ \frac{\tau_{B}}{e(1-\tau_{B})}\right] ,
\end{equation}
which goes to zero only for $\tau_{B}\rightarrow0$, corresponding
to Bob being arbitrarily far from the relay.

We can also explain why the other asymmetric configuration with
$\tau _{B}\simeq1$ is not particularly profitable (at fixed loss
rate this corresponds to Bob approaching the relay). Taking the
limit $\tau
_{B}\rightarrow1$ in Eq.~(\ref{Rate_tA_tB_chi}), we find%
\begin{align}
R_{\tau_{B}\rightarrow1} & =h\left[
\tau_{A}+(1-\tau_{A})\omega_{A}\right]
-h(\omega_{A})  \label{RateDR} \\
& +\log_{2}\left\{
\frac{2\tau_{A}}{e(1-\tau_{A})[1+\tau_{A}+(1-\tau
_{A})\omega_{A}]}\right\} ,  \notag
\end{align}
which depends on $\tau_{A}$ and $\omega_{A}$ only, as expected,
since the attack must reduce to an entangling cloner attack versus
Alice's link. Here it is important to note that Eq.~(\ref{RateDR})
coincides with the
direct-reconciliation rate of the point-to-point no-switching protocol~\cite%
{NoSwitchapp} (whose asymptotic analytical expression can be found
in the Supplementary Information of
Ref.~\cite{TwowayPROTOCOLapp}). As we know this
rate has a limited range. For instance, in the presence of pure loss ($%
\omega_{A}=1$), we have%
\begin{equation}
R_{\tau_{B}\rightarrow1}^{\text{loss}}=\log_{2}\left[ \frac{\tau_{A}}{%
e(1-\tau_{A})}\right] ,   \label{Rloss2}
\end{equation}
which is zero at $\tau_{A}\simeq0.73$, therefore limiting Alice's
distance from the relay to about $6.8$km in standard optical
fibres ($0.2$dB/km).

It is clear that the symmetric configuration
$\tau_{A}=\tau_{B}:=\tau$ has a performance which must be somehow
intermediate between the previous limit cases. More precisely the
performance of the symmetric configuration is comparable with that
of Bob approaching the relay. In fact, for pure-loss
attacks ($\omega_{A}=\omega_{B}=1$), we have the rate in Eq.~(\ref%
{pLOSSrateSYM}), which goes to zero for $\tau\simeq0.84$,
therefore restricting Alice's and Bob's distances from the relay
to about $3.8$km in optical fibres ($0.2$dB/km). This means that
the overall distance between Alice and Bob cannot exceed $7.6$km
when the relay is perfectly in the middle.

Finally, it is interesting to note how the Bell detection reverses
the role of\ the two types of reconciliations. In fact, suppose to
have an EPR source very close to Alice, who heterodynes one mode
in order to encode the signal variable in the other mode being
sent to Bob who is far away. In this scheme, Bob guessing Alice
corresponds to direct reconciliation. If we now replace the EPR
source with a relay performing a Bell detection on incoming modes,
the situation is reversed and Bob guessing Alice becomes
equivalent to reverse reconciliation.

\subsection{Computation of the post-relay CM\label{postCMapp}}

In this technical section we explicitly compute the formula of Eq.~(\ref%
{VabGamma}) for the post-relay CM $\mathbf{V}_{ab|\gamma}$
corresponding to the scenario depicted in Fig.~\ref{picAPP}. At
the input, Alice's modes $a$ and $A$, Bob's modes $b$ and $B$, and
Eve's modes $E_{1}$ and $E_{2}$ are in
a tensor-product state $\rho_{aA}\otimes\rho_{bB}\otimes\sigma_{E_{1}E_{2}}$%
, where $\rho_{aA}=\rho_{bB}=\rho$ is an EPR state with CM%
\begin{equation}
\mathbf{V}(\mu)=\left(
\begin{array}{cc}
\mu\mathbf{I} & \mu^{\prime}\mathbf{Z} \\
\mu^{\prime}\mathbf{Z} & \mu\mathbf{I}%
\end{array}
\right) ,~%
\begin{array}{c}
\mu^{\prime}:=\sqrt{\mu^{2}-1},~~~ \\
\mathbf{Z}:=\mathrm{diag}(1,-1),%
\end{array}%
\end{equation}
and $\sigma_{E_{1}E_{2}}$ is Eve's zero-mean Gaussian state with CM $\mathbf{%
V}_{E_{1}E_{2}}$ in the normal form of Eq.~(\ref{CMeveAPP}). The
global
state is then a zero-mean Gaussian state with CM%
\begin{equation}
\mathbf{V}_{aAbBE_{1}E_{2}}=\mathbf{V}(\mu)\oplus\mathbf{V}(\mu)\oplus
\mathbf{V}_{E_{1}E_{2}}.
\end{equation}

It is helpful to permute the modes so as to have the ordering
$abAE_{1}E_{2}B $, where the upper-case modes are those
transformed by the beam splitters.
After reordering, the input CM\ has the explicit form%
\begin{equation}
\mathbf{V}_{abAE_{1}E_{2}B}=\left(
\begin{array}{cccccc}
\mu\mathbf{I} & \mathbf{0} & \mu^{\prime}\mathbf{Z} & \mathbf{0} &
\mathbf{0}
& \mathbf{0} \\
\mathbf{0} & \mu\mathbf{I} & \mathbf{0} & \mathbf{0} & \mathbf{0}
&
\mu^{\prime}\mathbf{Z} \\
\mu^{\prime}\mathbf{Z} & \mathbf{0} & \mu\mathbf{I} & \mathbf{0} &
\mathbf{0}
& \mathbf{0} \\
\mathbf{0} & \mathbf{0} & \mathbf{0} & \omega_{A}\mathbf{I} &
\mathbf{G} &
\mathbf{0} \\
\mathbf{0} & \mathbf{0} & \mathbf{0} & \mathbf{G} &
\omega_{B}\mathbf{I} &
\mathbf{0} \\
\mathbf{0} & \mu^{\prime}\mathbf{Z} & \mathbf{0} & \mathbf{0} &
\mathbf{0} &
\mu\mathbf{I}%
\end{array}
\right) ,
\end{equation}
where $\mathbf{0}$ is the $2\times2$ zero matrix. Now the action
of the two beam splitters with transmissivities $\tau_{A}$ and
$\tau_{B}$ is described
by the symplectic matrix%
\begin{equation}
\mathbf{S=\mathbf{I}}\oplus\mathbf{\mathbf{I}}\oplus\mathbf{S\mathbf{(}}%
\tau_{A}\mathbf{\mathbf{)}\oplus S(}\tau_{B}\mathbf{)}^{T}~,
\label{SK}
\end{equation}
where the identity matrices
$\mathbf{\mathbf{I}}\oplus\mathbf{\mathbf{I}}$
act on the remote modes $a$ and $b$, the beam splitter matrix%
\begin{equation}
\mathbf{S}(\tau_{A})=\left(
\begin{array}{cc}
\sqrt{\tau_{A}}\mathbf{I} & \sqrt{1-\tau_{A}}\mathbf{I} \\
-\sqrt{1-\tau_{A}}\mathbf{I} & \sqrt{\tau_{A}}\mathbf{I}%
\end{array}
\right)   \label{BSmatrix}
\end{equation}
acts on modes $A$ and $E_{1}$, and the transposed beam splitter matrix $%
\mathbf{S}(\tau_{B})^{T}$ acts on modes $E_{2}$ and $B$.

The state describing the output modes $abA^{\prime}E_{1}^{\prime}E_{2}^{%
\prime}B^{\prime}$ after the beam splitters is a Gaussian state
with zero
mean and CM equal to%
\begin{equation}
\mathbf{V}_{abA^{\prime}E_{1}^{\prime}E_{2}^{\prime}B^{\prime}}=\mathbf{S~V}%
_{abAE_{1}E_{2}B}~\mathbf{S}^{T}~.
\end{equation}
After some algebra, we get%
\begin{equation}
\mathbf{V}_{abA^{\prime}E_{1}^{\prime}E_{2}^{\prime}B^{\prime}}=\left(
\begin{array}{ccc}
\mathbf{V}_{ab} & \mathbf{W}_{1} & \mathbf{W}_{2} \\
\mathbf{W}_{1}^{T} & \mathbf{V}_{A^{\prime}E_{1}^{\prime}} &
\mathbf{W}_{3}
\\
\mathbf{W}_{2}^{T} & \mathbf{W}_{3}^{T} & \mathbf{V}_{E_{2}^{\prime}B^{%
\prime }}%
\end{array}
\right) ,   \label{V_tot}
\end{equation}
where the blocks along the diagonal correspond to the reduced CMs $\mathbf{V}%
_{ab}=\mu(\mathbf{\mathbf{I}}\oplus\mathbf{\mathbf{I}})$,
\begin{equation}
\mathbf{V}_{A^{\prime}E_{1}^{\prime}}=\left(
\begin{array}{cc}
x_{A}\mathbf{I} & x_{A}^{\prime\prime}\mathbf{I} \\
x_{A}^{\prime\prime}\mathbf{I} & x_{A}^{\prime}\mathbf{I}%
\end{array}
\right) ,~\mathbf{V}_{E_{2}^{\prime}B^{\prime}}=\left(
\begin{array}{cc}
x_{B}^{\prime}\mathbf{I} & x_{B}^{\prime\prime}\mathbf{I} \\
x_{B}^{\prime\prime}\mathbf{I} & x_{B}\mathbf{I}%
\end{array}
\right) ,
\end{equation}
where we have set (for $k=A,B$)%
\begin{align}
x_{k} & :=\tau_{k}\mu+(1-\tau_{k})\omega_{k}, \\
x_{k}^{\prime} & :=\tau_{k}\omega_{k}+(1-\tau_{k})\mu, \\
x_{k}^{\prime\prime} &
:=\sqrt{\tau_{k}(1-\tau_{k})}(\omega_{k}-\mu).
\end{align}
The off-diagonal blocks are given by%
\begin{equation}
\mathbf{W}_{1}=\left(
\begin{array}{cc}
\mu^{\prime}\sqrt{\tau_{A}}\mathbf{Z} & -\mu^{\prime}\sqrt{1-\tau_{A}}%
\mathbf{Z} \\
\mathbf{0} & \mathbf{0}%
\end{array}
\right) ,
\end{equation}%
\begin{equation}
\mathbf{W}_{2}=\left(
\begin{array}{cc}
\mathbf{0} & \mathbf{0} \\
-\mu^{\prime}\sqrt{1-\tau_{B}}\mathbf{Z} & \mu^{\prime}\sqrt{\tau_{B}}%
\mathbf{Z}%
\end{array}
\right) ,
\end{equation}
and%
\begin{equation}
\mathbf{W}_{3}=\left(
\begin{array}{cc}
\sqrt{(1-\tau_{A})\tau_{B}}\mathbf{G} & \sqrt{(1-\tau_{A})(1-\tau_{B})}%
\mathbf{G} \\
\sqrt{\tau_{A}\tau_{B}}\mathbf{G} & \sqrt{\tau_{A}(1-\tau_{B})}\mathbf{G}%
\end{array}
\right) .
\end{equation}

Since we are interested in the output CM\ of Alice and Bob, we trace out $%
E_{1}^{\prime}$ and $E_{2}^{\prime}$, which corresponds to
deleting the corresponding rows and columns in the CM of
Eq.~(\ref{V_tot}). As a result,
we get the following reduced CM for modes $abA^{\prime}B^{\prime}$%
\begin{equation}
\mathbf{V}_{abA^{\prime}B^{\prime}}=\left(
\begin{array}{ccc}
\mathbf{V}_{ab} & \mathbf{C}_{1} & \mathbf{C}_{2} \\
\mathbf{C}_{1}^{T} & \mathbf{A} & \mathbf{D} \\
\mathbf{C}_{2}^{T} & \mathbf{D}^{T} & \mathbf{B}%
\end{array}
\right) ,   \label{V_seconda}
\end{equation}
where the various blocks are given by%
\begin{align}
\mathbf{A} & =x_{A}\mathbf{I,~B}=x_{B}\mathbf{I,} \\
\mathbf{D} & =\sqrt{(1-\tau_{A})(1-\tau_{B})}\mathbf{G,}
\end{align}
and%
\begin{equation}
\mathbf{C}_{1}=\left(
\begin{array}{c}
\mu^{\prime}\sqrt{\tau_{A}}\mathbf{Z} \\
\mathbf{0}%
\end{array}
\right) ,~\mathbf{C}_{2}=\left(
\begin{array}{c}
\mathbf{0} \\
\mu^{\prime}\sqrt{\tau_{B}}\mathbf{Z}%
\end{array}
\right) .
\end{equation}

From the CM of Eq.~(\ref{V_seconda}) we derive the CM
$\mathbf{V}_{ab|\gamma} $ of the conditional remote state
$\rho_{ab|\gamma}$ after the Bell measurement on modes
$A^{\prime}$ and $B^{\prime}$. For this derivation, we use the
transformation rules for CMs under Bell-like measurements given in
Ref.~\cite{BellFORMULAapp}. From the blocks of the
CM~(\ref{V_seconda}), we
construct the following theta matrix%
\begin{equation}
\boldsymbol{\Theta}:=\frac{1}{2}\left( \mathbf{ZAZ}+\mathbf{B}-\mathbf{ZD}-%
\mathbf{D}^{T}\mathbf{Z}\right) .   \label{stGAMMA}
\end{equation}
After some simple algebra, we find
\begin{equation}
\boldsymbol{\Theta}=\frac{1}{2}\left(
\begin{array}{cc}
\theta & 0 \\
0 & \theta^{\prime}%
\end{array}
\right) ,
\end{equation}
whose diagonal elements are given in Eq.~(\ref{teta2}).

Then the conditional CM is given by the
formula~\cite{BellFORMULAapp}
\begin{equation}
\mathbf{V}_{ab|\gamma}=\mathbf{V}_{ab}-\frac{1}{2\det\boldsymbol{\Theta}}%
\sum_{i,j=1}^{2}\mathbf{C}_{i}(\mathbf{X}_{i}^{T}\boldsymbol{\Theta}\mathbf{X%
}_{j})\mathbf{C}_{j}^{T},   \label{stBELL}
\end{equation}
where%
\begin{equation}
\mathbf{X}_{1}:=\left(
\begin{array}{cc}
0 & 1 \\
1 & 0%
\end{array}
\right) ,~\mathbf{X}_{2}:=\left(
\begin{array}{cc}
0 & 1 \\
-1 & 0%
\end{array}
\right) .   \label{XeOM}
\end{equation}
After some algebra, we find the expression in
Eq.~(\ref{VabGamma}).

\section{Experimental methods\label{EXP_app}}

In this section we start by providing a general description of the
experimental setup, discussing the main optical elements involved
in the implementation (Sec.~\ref{setup_APP}). Then, in
Sec.~\ref{data_APP}, we give a more detailed mathematical
interpretation of the experiment, and we discuss the data
post-processing together with the finite-size effects associated
with the protocol.

\subsection{General description of the optical setup\label{setup_APP}}

In our proof-of-principle experiment we used a highly stable laser at $1064$%
nm which is split in equal portions and directed to the stations
of Alice and Bob, providing them with a common local oscillator.
In a future in-field implementation of the protocol, such a
phase-locking of the beams can be achieved by atom-clock
syncronization and classical communication between the parties. In
general, the local oscillator could even be provided by the
eavesdropper since does not contain any information about the
encodings of Alice and Bob. Furthermore, it can be continuously
monitored by the parties and suitably filtered to delete the
presence of additional degrees of freedom (side-channel
attacks~\cite{SidePRLapp}).

\begin{figure}[ptbh]
\vspace{-0.0cm}
\par
\begin{center}
\includegraphics[width=0.48\textwidth] {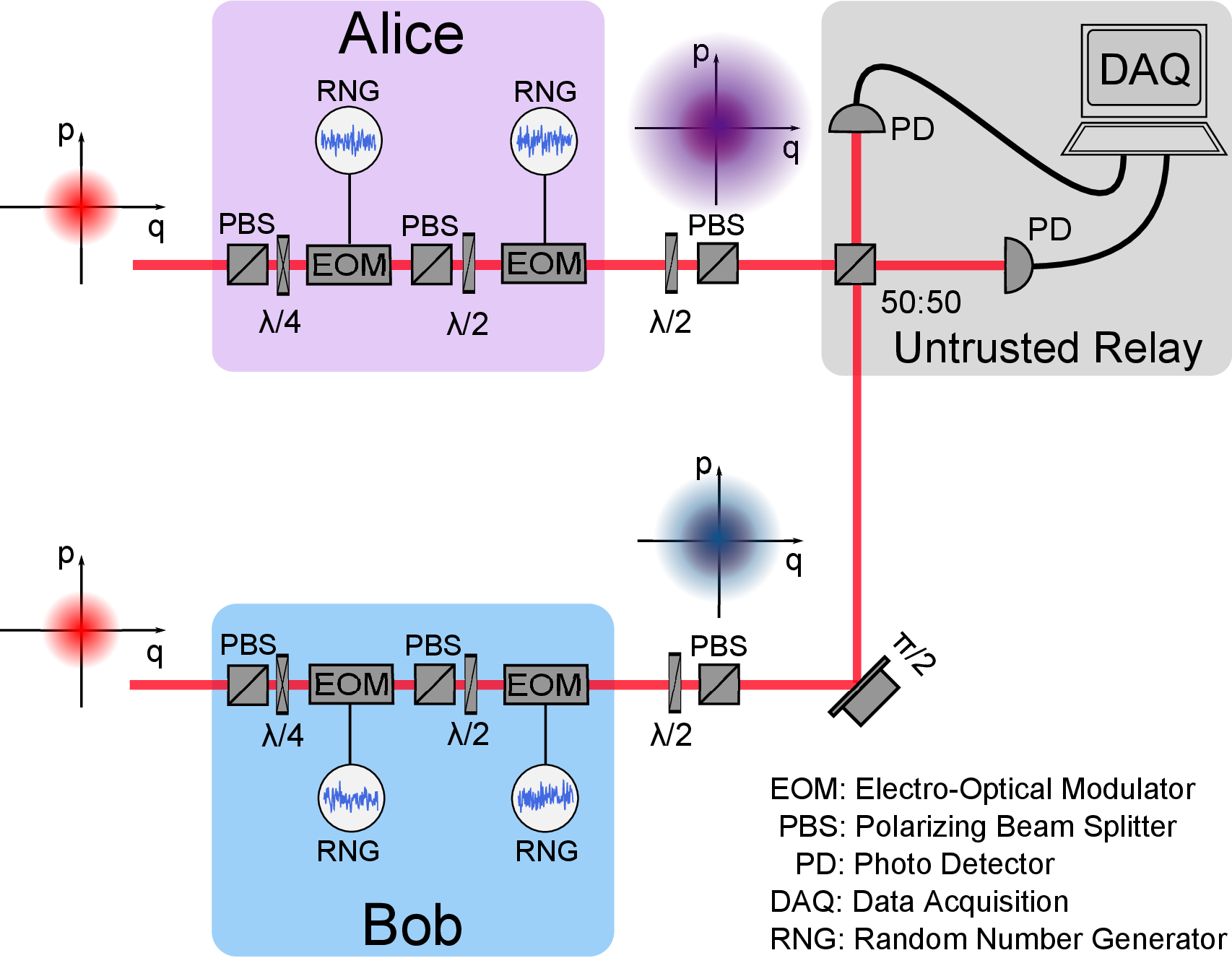}
\end{center}
\par
\vspace{-0.3cm} \caption{Experimental setup (same as Fig.~5 in the
main text). Alice and Bob apply amplitude and phase modulators to
a pair of identical classical phase-locked bright coherent beams,
providing a common local oscillator and coming from a highly
stable laser (not shown). At the output, the two modes emerge
randomly-displaced in the phase space according to a Gaussian
distribution. In particular, Bob's modulation is suitably
attenuated to simulate loss in his link. Bob's mode is then
phase-shifted and merged with Alice's mode at the relay's balanced
beam splitter. The two output ports are photodetected and
processed to realize an equivalent continuous-variable Bell
measurement.} \label{expPIC}
\end{figure}

At each of the stations (see Fig.~\ref{expPIC}), the laser beams
are modulated using amplitude and phase electro-optical modulators
that are fed by Gaussian modulated signals from independent
electronic signal generators. The Gaussian modulations are white
within the measurement bandwidth. Prior to modulation, the laser
fields were polarized to a very high degree to ensure pure
amplitude (phase) modulation produced by the amplitude (phase)
modulator. Furthermore, the transverse profiles of the laser beams
were made as large as possible in the modulator to ensure
preservation of their Gaussian profiles.

At the relay, the two beams interfere at a balanced beam splitter
with a visibility of $96\%$ and their relative phase is actively
controlled with a piezo mounted mirror to produce equally intense
output beams. These beams are focused onto two balanced detectors
and the resulting currents are subtracted and added in order to
produce the difference of the amplitude quadratures and the sum of
the phase quadratures, respectively. This method is a simple
alternative to the standard eight-port measurement setup needed
for the continuous-variable Bell detection and is enabled by the
brightness of the carrier~\cite{UlrikPRL}. As the subtraction and
addition processes are performed in a software program, an
imbalanced hardware-system can be compensated during the
post-processing (See the next section for more details).

The measurements are carried out at the sideband frequency of
$10.5$MHz which is well separated from the carrier frequency
thereby avoiding low frequency noise and thus ensuring quantum
noise limited performance of the protocol. The power of the
individual laser beams was $1.4$mW. The signal is mixed down to DC
from $10.5$MHz, low pass filtered at $100$kHz and digitized with a
sampling rate of $500$kHz and $14$bit resolution. Each data block
consists of $10^{6}$ data points corresponding to a measurement
time of $2$ seconds.

The modulation depth in Bob's link was reduced to simulate the
transmission loss. A reduction in modulation is equivalent to a
beam splitter induced transmission loss because both reduce the
power in the sidebands. The difference is that a reduction in
modulation depth keeps the power of the carrier (with respect to
the sidebands) laser beam constant, which was experimentally
convenient. This allows us to accurately simulate the transmission
through the channels and thus test different communication
realizations (see the next section for more details on this
method).

%In order to provide a visual idea on how the Bell relay creates a posteriori
%correlations between Alice and Bob, see the example in
%Fig.~\ref{CorrelationsPIC} which refers to $10$dB loss in Bob's link. The two
%top panels in the figure describe the..... \begin{figure}[ptbh]
%\vspace{-0.0cm}
%\par
%\begin{center}
%\includegraphics[width=0.36\textwidth] {MDI-correlations.eps}
%\end{center}
%\par
%\vspace{-0.3cm}\caption{Schematic of the experimental setup. A bright laser
%beam (local oscillator) is split into equal parts and sent to Alice and Bob,
%which apply independent amplitude and phase modulators. Alice's mode is
%phase-shifted by $\pi/2$ before being merged with Bob's mode at the relay's
%balanced beamsplitter. The two outputs are photodetected and processed to
%simulate a continuous-variable Bell measurement.}%
%\label{CorrelationsPIC}%
%\end{figure}

\subsection{Detailed mathematical description of the experiment and data
post-processing\label{data_APP}}

Consider the schematic of Fig.~\ref{expPIC}. The two input
classical beams are identical and phase-locked, each providing a
bright coherent state with amplitude $iL$ ($L\gg1$). Note that we
have complete freedom in choosing the global phase of this
oscillator, which is here set to $\pi$ just for convenience of
notation.

At the output of the modulators, Alice's and Bob's annihilation
operators can be written as $\hat{A}=iL+\hat{a}$ and
$\hat{B}=iL+\hat{b}$, respectively. Before entering the second
balanced beam splitter at the relay, Bob's mode is phase-shifted
by $\pi/2$, so that the two output ports have roughly the same
intensity. This is equivalent to transform Alice's mode as
$\hat{A}\rightarrow i\hat{A}$ (since this is a relative shift, it
can be mathematically applied to Alice or Bob). The output ports
of the beam splitter are then described by the following
annihilation operators
\begin{equation}
\hat{D}_{0}=\frac{i\hat{A}+\hat{B}}{\sqrt{2}},~\hat{D}_{1}=\frac{i\hat{A}-%
\hat{B}}{\sqrt{2}}.
\end{equation}

Each output port ($k=0,1$) is measured by a photodetector with photocurrent $%
i_{k}=c_{k}\hat{N}_{k}$, where
$\hat{N}_{k}=\hat{D}_{k}^{\dagger}\hat{D}_{k}$ is the number
operator, and $c_{k}$ is an optical-to-current conversion factor.
This factor is different for the two detectors, so that they have
different levels of electronic shot-noise. To counterbalance this
asymmetry, we can re-scale one of the currents by a real factor
$g$. Thus, by
re-scaling the currents and taking their sum and difference, we get%
\begin{equation}
\Sigma_{\pm}:=i_{0}\pm g~i_{1}=c_{0}\left( \hat{N}_{0}\pm r\hat{N}%
_{1}\right) ,
\end{equation}
where $r:=gc_{1}/c_{0}$. By expanding at the first order in the
local oscillator $L$, we derive
\begin{align}
\frac{\Sigma_{+}}{c_{0}} & =(1+r)L^{2}+\frac{L}{2}\left[ (1-r)\hat{q}%
_{-}+(1+r)\hat{p}_{+}\right] , \\
\frac{\Sigma_{-}}{c_{0}} & =(1-r)L^{2}+\frac{L}{2}\left[ (1+r)\hat{q}%
_{-}+(1-r)\hat{p}_{+}\right] .
\end{align}

The next step is subtracting the offset of the local oscillator
which is equivalent to subtracting the mean values $\left\langle
\Sigma_{\pm }\right\rangle =c_{0}(1\pm r)L^{2}$. The result is
normalized dividing by the standard deviation of the vacuum
fluctuations
\begin{equation}
\sigma_{\pm}=\sqrt{\left\langle \Sigma_{\pm}^{2}\right\rangle _{\text{vac}%
}-\left\langle \Sigma_{\pm}\right\rangle _{\text{vac}}^{2}}=c_{0}L\sqrt {%
\frac{1+r^{2}}{2}}.
\end{equation}
Thus, we get
\begin{equation}
\frac{\Sigma_{+}-\left\langle \Sigma_{+}\right\rangle }{\sigma_{+}}=\hat {x}%
_{+r},~~\frac{\Sigma_{-}-\left\langle \Sigma_{-}\right\rangle }{\sigma_{-}}=%
\hat{x}_{-r},
\end{equation}
where
\begin{equation}
\hat{x}_{+r}:=\kappa_{1}\hat{q}_{-}+\kappa_{2}\hat{p}_{+}
\label{linCOMB}
\end{equation}
is a linear combination $\hat{q}_{-}$ and $\hat{p}_{+}$, with coefficients%
\begin{equation}
\kappa_{1}=\frac{1-r}{\sqrt{2(1+r^{2})}},~\kappa_{2}=\frac{1+r}{\sqrt {%
2(1+r^{2})}}.
\end{equation}

We can rewrite the linear combination of Eq.~(\ref{linCOMB}) as%
\begin{equation}
\hat{x}_{+r}=\frac{1-r}{1+r}~\kappa_{2}\hat{q}_{-}+\kappa_{2}\hat{p}_{+},
\end{equation}
and include the factor $\kappa_{2}$ in Alice's and Bob's classical
modulations. In fact, the quantum variables can always be decomposed as%
\begin{equation}
\hat{q}_{-}=q_{-}+\hat{q}_{-}^{\text{vac}},~\hat{p}_{+}=p_{+}+\hat{p}_{+}^{%
\text{vac}},
\end{equation}
where $q_{-}$ and $p_{+}$ are the classical parts, with $\hat{q}_{-}^{\text{%
vac}}$ and $\hat{p}_{+}^{\text{vac}}$ accounting for vacuum noise.
Then, we can write
\begin{equation}
\hat{x}_{+r}=\frac{1-r}{1+r}q_{-}+p_{+}+\hat{\delta}_{+}^{\text{vac}},
\label{xg}
\end{equation}
where $\hat{\delta}_{+}^{\text{vac}}$ is vacuum noise and the
classical variables have been re-scaled as
\begin{equation}
\kappa_{2}q_{-}\rightarrow q_{-},~\kappa_{2}p_{+}\rightarrow
p_{+}.
\end{equation}
This is equivalent to re-scaling Alice's and Bob's classical variables%
\begin{equation}
\kappa_{2}\left(
\begin{array}{c}
q_{A} \\
p_{A} \\
q_{B} \\
p_{B}%
\end{array}
\right) \rightarrow\left(
\begin{array}{c}
q_{A} \\
p_{A} \\
q_{B} \\
p_{B}%
\end{array}
\right) ,
\end{equation}
so that their new variables are Gaussianly-modulated with a
re-scaled
variance $\kappa_{2}^{2}\varphi\rightarrow\varphi$. Similarly to Eq.~(\ref%
{xg}), we derive%
\begin{equation}
\hat{x}_{-r}=q_{-}+\frac{1-r}{1+r}p_{+}+\hat{\delta}_{-}^{\text{vac}}.
\label{pg}
\end{equation}

Thus, the outcome of the relay is generally given by the pair $%
(x_{-r},x_{+r})$ or equivalently
$\gamma_{r}:=(x_{-r}+ix_{+r})/\sqrt{2}$, where $r$ is a parameter
which can be optimized. Note that, for $r=1$, we have $\hat
{x}_{+1}=\hat{p}_{+}$ and $\hat{x}_{-1}=\hat{q}_{-}$. In general,
for the presence of quadrature asymmetries and $q$-$p$
correlations (coming from the cross-talk between the amplitude and
phase modulators), the optimal value of $r$ may be different from
$1$, e.g., in the range $0.45\div0.75$ in our experiment. This
optimization is a simple operation which enables us to
counterbalance some of the technical imperfections in our setup.
Similar optimizations could also be exploited in potential
in-field implementations of the protocol (given an observed
statistics, Alice and Bob can always assume an optimized relay and
ascribe all the noise to the coherent attack of the links.)

In order to use Eqs.~(\ref{xg}) and~(\ref{pg}), we need to access
the
experimental values of Alice's and Bob's optical displacements $%
(q_{A},p_{A},q_{B},p_{B})^{T}$. Therefore we have to compute the
electro-optical gains of the modulators, for both amplitude and
phase (i.e.,
position and momentum on top of the bright local oscillator). These gains $%
\{t_{1},t_{2},t_{3},t_{4}\}$ convert the applied electronic displacements $\{%
\mathcal{A}_{q},\mathcal{B}_{q},\mathcal{A}_{p},\mathcal{B}_{p}\}$
into the optical quadrature displacements. For their computation,
we minimize the following variances
\begin{align}
& \left\langle \left[ \hat{x}_{-r}-\frac{t_{1}\mathcal{A}_{q}-t_{2}\mathcal{B%
}_{q}}{\sqrt{2}}\right] ^{2}\right\rangle , \\
& \left\langle \left[ \hat{x}_{+r}-\frac{t_{3}\mathcal{A}_{p}+t_{4}\mathcal{B%
}_{p}}{\sqrt{2}}\right] ^{2}\right\rangle .
\end{align}
Once these gains are known, we can experimentally determine the
optical displacements of Alice and Bob as
\begin{align}
q_{A} & =t_{1}\mathcal{A}_{q},~q_{B}=t_{2}\mathcal{B}_{q}, \\
p_{A} & =t_{3}\mathcal{A}_{p},~p_{B}=t_{4}\mathcal{B}_{p}.
\end{align}

The classical modulations in the optical quadratures are
approximately equal, with a maximal variance of $\varphi\simeq65$
vacuum-noise units. In order to simulate beam splitters in the
links, we equivalently attenuate the
classical modulations $\varphi_{A}=V(q_{A})\simeq V(p_{A})$ and $%
\varphi_{B}=V(q_{B})\simeq V(p_{B})$. In fact, an ensemble of
Gaussian-modulated coherent states is described by an average
state which is thermal with quantum variance $\mu=\varphi+1$,
where $\varphi$ is the classical modulation and $1$ is the
variance of the vacuum noise. By sending a thermal state with such
a variance through a beam splitter with
transmissivity $\tau$ (in a vacuum environment), we get the output variance%
\begin{equation}
V=\tau(\varphi+1)+1-\tau=\tau\varphi+1.
\end{equation}
This is equivalent to removing the beam splitter and sending a
thermal state with quantum variance $\tau\varphi+1$, i.e.,
Gaussian-modulated coherent states with a reduced modulation
$\tau\varphi$.

Thus, Bob's modulation is taken to be $\varphi_{B}=\tau_{B}\varphi$ with $%
\tau_{B}$ being the equivalent transmissivity of the beam
splitter. We have considered several values of $\tau_{B}$ down to
about $4\times10^{-4}$,
corresponding to $34$dB loss (equivalent to $170$km in optical fibre at $0.2$%
dB/km). Alice's modulation is fixed to be maximal $\varphi_{A}\simeq\varphi$%
, simulating the scenario where Alice's link has $\tau_{A}\simeq1$
(small effects due to non-unit quantum efficiencies of the
detectors and non-unit beam splitter visibility have been
neglected by suitably re-scaling Alice's modulation). Note that,
in a future real-time implementation of the protocol, Alice's
transmissivity can indeed be very close to $1$ by using
photodiodes with high quantum efficiencies ($\sim$98\%).

Thus, in our experiment we realize the conditions%
\begin{align}
\hat{x}_{-r} & \simeq\frac{q_{A}-\sqrt{\tau_{B}}q_{B}}{\sqrt{2}}+\frac {1-r}{%
1+r}\frac{p_{A}+\sqrt{\tau_{B}}p_{B}}{\sqrt{2}}+\hat{\delta}_{-}^{\text{vac}%
},  \label{x-r} \\
\hat{x}_{+r} & \simeq\frac{1-r}{1+r}\frac{q_{A}-\sqrt{\tau_{B}}q_{B}}{\sqrt{2%
}}+\frac{p_{A}+\sqrt{\tau_{B}}p_{B}}{\sqrt{2}}+\hat{\delta}_{+}^{\text{vac}%
},   \label{x+r}
\end{align}
which connect Alice's amplitude $\alpha:=(q_{A}+ip_{A})/2$, Bob's amplitude $%
\beta:=(q_{B}+ip_{B})/2$, and the relay outcome $%
\gamma_{r}:=(x_{-r}+ix_{+r})/\sqrt{2}$ under a pure-loss attack of
Bob's link. One can easily check that, for $r=1$, the previous
Eqs.~(\ref{x-r})
and~(\ref{x+r}) become%
\begin{align}
\hat{q}_{-} & \simeq\frac{q_{A}-\sqrt{\tau_{B}}q_{B}}{\sqrt{2}}+\hat{\delta }%
_{-}^{\text{vac}}, \\
\hat{p}_{+} & \simeq\frac{p_{A}+\sqrt{\tau_{B}}p_{B}}{\sqrt{2}}+\hat{\delta }%
_{+}^{\text{vac}}.
\end{align}

For any experimental value of $\tau_{B}$, we have considered an
optimized
relay (with optimal $r$ published) and collected $10^{6}$ values of $\alpha$%
, $\beta$ and $\gamma_{r}$. The value of $\tau_{B}$ is accessible
to Alice and Bob by computing and comparing the first moments of
their data which must follow Eqs.~(\ref{x-r}) and~(\ref{x+r}).
Their estimate is in good agreement with the applied attenuation.
The parties can also derive the global classical CM
$\mathbf{V}(q_{A},p_{A},q_{B},p_{B},x_{-r},x_{+r})$ by comparing a
subset of their data. Once this is known, they can derive the
secret key rate of the protocol.

\begin{figure}[ptbh]
\vspace{-0.4cm}
\par
\begin{center}
\includegraphics[width=0.46\textwidth] {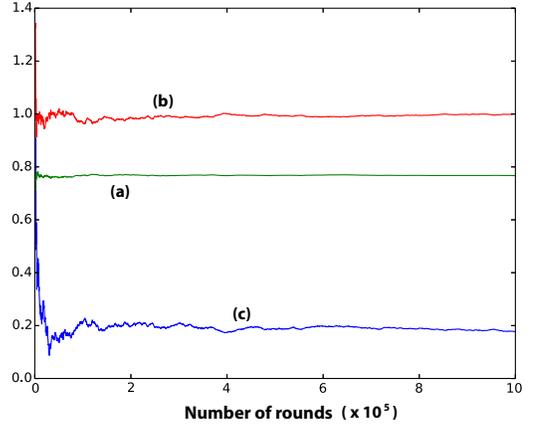}
\end{center}
\par
\vspace{-0.5cm} \caption{As the number of rounds of the protocol
increases up to $10^{6}$, we plot Alice and Bob's estimate of the
transmissivity $\protect\tau_{B}$ (a), the estimate of the scaled
determinant of the global CM\ $\mathbf{V}$ (b) and the estimate of
the key rate of the protocol (c). Asymptotic values are quickly
reached after about $10^{5}$ experimental points. This behaviour
is generic in $r$ (here we have chosen $r=1$) and for the various
experimental points (here we have chosen
$\protect\tau_{B}\simeq0.77$).} \label{sizePIC}
\end{figure}

Before proceeding, let us first discuss the experimental
finite-size effects involved in the estimate of the first- and
second-order moments. As we can see from Fig.~\ref{sizePIC}, the
asymptotic values of the statistical moments are already reached
after $\simeq10^{5}$ rounds of the protocol (for the second-order
moments we have plotted the scaled determinant of the global CM,
i.e., $\det\mathbf{V}_{n}/\det\mathbf{V}_{\infty}$, where $n$ is
the variable number of rounds and $\infty=10^{6}$). Thus, the
parties rapidly reach the asymptotic regime, where the finite-size
effects on the key rate can be neglected (see again
Fig.~\ref{sizePIC} for the rate). Thus the parties just need to
compare $\simeq10^{5}$ points in order to estimate the asymptotic
values of the statistical moments. Clearly, this represents a
negligible subset of data in a real-time implementation of the
protocol
where the number of rounds can be $\gg10^{5}$ (e.g., $>10^{9}$ in Ref.~\cite%
{Grangier2app}).

Let us now derive the experimental key rate from the global CM\ $\mathbf{V}%
(q_{A},p_{A},q_{B},p_{B},x_{-r},x_{+r})$. The first step is to
perform the Gaussian elimination of the relay variables in order
to construct the
conditional CM\ $\mathbf{V}_{\text{cond}}:=\mathbf{V}%
(q_{A},p_{A},q_{B},p_{B}|\gamma_{r})$ according to Eqs.~(\ref{decogCM}) and~(%
\ref{GaussELI}). This conditional matrix has still some
asymmetries in the quadratures plus residual $q$-$p$ correlations.
These imperfections can be counterbalanced by Alice and Bob by
performing local rotations and re-scalings of their classical data
$\alpha$ and $\beta$. By means of these local symplectic
transformations, the conditional matrix is transformed into a
normal form
\begin{equation}
\mathbf{V}_{\text{cond}}=\left(
\begin{array}{cc}
\mathbf{A} & \mathbf{C} \\
\mathbf{C}^{T} & \mathbf{B}%
\end{array}
\right) \rightarrow\left(
\begin{array}{cc}
a\mathbf{I} & c\mathbf{Z} \\
c\mathbf{Z} & b\mathbf{I}%
\end{array}
\right) ,
\end{equation}
where $a=\sqrt{\det\mathbf{A}}$, $b=\sqrt{\det\mathbf{B}}$ and $c$
is determined by computing the other two symplectic invariants
$\det\mathbf{V}$ and
$\det\mathbf{A}+\det\mathbf{B}+2\det\mathbf{C}$.

Once Alice and Bob have symmetrized the conditional CM, they
derive the quantum CM $\mathbf{V}_{ab|\gamma_{r}}$ in the
equivalent entanglement-based
representation of the protocol. According to Eq.~(\ref{V1appEX}) and~(\ref%
{classCMoutcomes}), this is given by%
\begin{equation}
\mathbf{V}_{ab|\gamma_{r}}=\eta^{2}\mathbf{V}_{\text{cond}}-\mathbf{I},
\end{equation}
where $\eta$ is the parameter of Eq.~(\ref{alphaTILD}) and
accounts for the finite modulation ($\eta\simeq1.01$ in the
experiment, where $\eta=1$ is the asymptotic regime of infinite
modulation).

From $\mathbf{V}_{ab|\gamma_{r}}$ we can easily extract the reduced CM $%
\mathbf{V}_{b|\gamma_{r}}$\ and compute the doubly-conditional CM $\mathbf{V}%
_{b|\gamma_{r}\tilde{\alpha}}$ via Eq.~(\ref{ff1}) or~(\ref{ff2}).
These matrices provide Alice and Bob's mutual information
$I_{AB}$\ according to Eqs.~(\ref{mutuaINab})
and~(\ref{sigmaAPP}). Then, we compute Eve's Holevo
information $I_{E}$ using Eq.~(\ref{HolINFOapp}), where the entropies of $%
\rho_{ab|\gamma_{r}}$ and $\rho_{b|\gamma_{r}\tilde{\alpha}}$ are
computed
from the symplectic spectra of $\mathbf{V}_{ab|\gamma_{r}}$ and $\mathbf{V}%
_{b|\gamma_{r}\tilde{\alpha}}$, respectively. Finally, we can
derive the experimental rate
\begin{equation}
R=\xi I_{AB}-I_{E},
\end{equation}
where $\xi$ is the reconciliation efficiency, with ideal value
$\xi=1$ and current achievable value of
$\xi\simeq97\%$~\cite{Jouguetapp}. The corresponding rates are
plotted in the main text and compared with the theoretical curves.
According to these final results, our adjustment of the
experimental imperfections\ is not perfect, with some residual
noise affecting our data. Such noise affects the rate in the same
way as a coherent Gaussian attack with excess noise
$\varepsilon\lesssim0.02$.


\begin{thebibliography}{99}                                                                                               %


\bibitem {Gisin}Gisin, N., Ribordy, G., Tittel, W. \& Zbinden, H. Quantum
cryptography. \textit{Rev. Mod. Phys.} \textbf{74}, 145 (2002).

\bibitem {Scarani}Scarani, V., Bechmann-Pasquinucci, H., Cerf, N. J., Dusek,
M., Lutkenhaus, N. \& Peev, M. The security of practical quantum
key distribution. \textit{Rev. Mod. Phys.} \textbf{81}, 1301
(2009).

\bibitem {Wilde}Wilde, M. M. \textit{Quantum Information Theory} (Cambridge
University Press, Cambridge, 2013).

\bibitem {RMP}Weedbrook, C., Pirandola, S., Garcia-Patron, R., Cerf, N. J.,
Ralph, T. C., Shapiro, J. H. \& Lloyd, S. Gaussian quantum
information. \textit{Rev. Mod. Phys.} \textbf{84}, 621 (2012).

\bibitem {SECOQC}SECOQC, 2007, http://www.secoqc.net.

\bibitem {SECOQC2}Peev, M. \textit{et al}. The SECOQC quantum key distribution
network in Vienna. \textit{New J. Phys.} \textbf{11}, 075001
(2009).

\bibitem {Tokyo1}Tokyo QKD network 2010, www.uqcc.org/QKDnetwork.

\bibitem {Tokyo2}Sasaki, M. \textit{et al.} Field test of quantum key
distribution in the Tokyo QKD Network. \textit{Opt. Express}
\textbf{19}, pp. 10387-10409 (2011).

\bibitem {SD4}Lydersen, L., Wiechers, C., Wittman, C., Elser, D., Skaar, D. \&
Makarov V. Hacking commercial quantum cryptography systems by
tailored bright illumination. \textit{Nature Photonics}
\textbf{4}, 686 (2010).

\bibitem {SD6}Gerhardt, I., Liu, Q., Lamas-Linares A., Skaar J., Kurtsiefer C.
\& Makarov, V. Full-field implementation of a perfect eavesdropper
on a quantum cryptography system. \textit{Nature Comm.}
\textbf{2}, 349 (2011).

\bibitem {SidePRL}Braunstein, S. L. \& Pirandola, S. Side-channel-free quantum
key distribution. \textit{Phys. Rev. Lett.} \textbf{108}, 130502
(2012).

\bibitem {Lo}Lo, H.-K., Curty, M. \& Qi, B. Measurement-device-independent
quantum key distribution. \textit{Phys. Rev. Lett.} \textbf{108},
130503 (2012).

\bibitem {Oth1}Ma, X., Fred Fung, C.-H. \& Razavi, M. Statistical fluctuation
analysis for measurement-device-independent quantum key
distribution. \textit{Phys. Rev. A} \textbf{86}, 052305 (2012).

\bibitem {Oth3}Ma, X. \& Razavi, M. Alternative schemes for
measurement-device-independent quantum key distribution.
\textit{Phys. Rev. A} \textbf{86}, 062319 (2012).

\bibitem {Others}Wang, X.B. Three-intensity decoy state method for device
independent quantum key distribution with basis dependent errors.
\textit{Phys. Rev. A} \textbf{87}, 012320 (2013).

\bibitem {Oth2}Branciard, C., Rosset, D., Liang, Y.-C. \& Gisin, N.
Measurement-device-independent entanglement witnesses for all
entangled quantum states. \textit{Phys. Rev. Lett.} \textbf{110},
060405 (2013).

\bibitem {Oth4}Tomamichel, M., Fehr, S., Kaniewski, J. \& Wehner, S. A
monogamy-of-entanglement game with applications to
device-independent quantum cryptography. \textit{New J. Phys.}
\textbf{15}, 103002 (2013).

\bibitem {Oth4b}Ci Wen Lim, C., Portmann, C., Tomamichel, M., Renner, R. \&
Gisin, N. Device-independent quantum key distribution with local
Bell test. \textit{Phys. Rev. X} \textbf{3}, 031006 (2013).

\bibitem {Oth5}Abruzzo, S., Kampermann, H., \& Bru\ss \ D.
Measurement-device-independent quantum key distribution with
quantum memories. \textit{Phys. Rev. A} \textbf{89}, 012301
(2014).

\bibitem {EXP1}Rubenok, A., Slater, J. A., Chan, P., Lucio-Martinez, I. \&
Tittel, W. Real-World two-photon interference and
proof-of-principle quantum key distribution immune to detector
attacks. \textit{Phys. Rev. Lett.} \textbf{111}, 130501 (2013).

\bibitem {EXP2}Ferreira da Silva, T., Vitoreti, D., Xavier, G. B., do Amaral,
G. C., Tempor\~{a}o, G. P. \& von der Weid, J. P.
Proof-of-principle demonstration of measurement-device-independent
quantum key distribution using polarization qubits. \textit{Phys.
Rev. A} \textbf{88}, 052303 (2013).

\bibitem {EXP3}Tang, Y.-L. \textit{et al.} Measurement-device-independent
quantum key distribution over 200 km. Preprint arxiv:1407.8012.

\bibitem {Nicolas2001}Cerf, N. J., Levy, M. \& van Assche, G. Quantum
distribution of Gaussian keys with squeezed states. \textit{Phys.
Rev. }\textbf{A} 63, 052311 (2001).

\bibitem {Grangier}Grosshans, F., van Assche, G., Wenger, J., Tualle-Brouri,
R., Cerf, N. J. \& Grangier, P. High-rate quantum cryptography
using Gaussian-modulated coherent states. \textit{Nature} (London)
\textbf{421}, 238 (2003).

\bibitem {NoSwitch}Weedbrook, C., Lance, A. M., Bowen, W. P., Symul, T.,
Ralph, T. C. \& Lam, P. K. Quantum cryptography without switching.
\textit{Phys. Rev. Lett.} \textbf{93}, 170504 (2004).

\bibitem {TwowayPROTOCOL}Pirandola, S., Mancini, S., Lloyd, S. \& Braunstein,
S. L. Continuous-variable quantum cryptography with two-way
quantum communication. \textit{Nature Phys.} \textbf{4}, 726
(2008).

\bibitem {Grangier2}Jouguet, P., Kunz-Jacques, S., Leverrier, A., Grangier, P.
\& Diamanti, E. Experimental demonstration of long-distance
continuous-variable quantum key distribution. \textit{Nature
Photonics} \textbf{7}, 378--381 (2013).

\bibitem {Ekert}Ekert, A., K. Quantum cryptography based on Bell's theorem.
\textit{Phys. Rev. Lett}. \textbf{67}, 661 (1991).

\bibitem {WeedEPR}Weedbrook, C. Continuous-variable quantum key distribution
with entanglement in the middle. \textit{Phys. Rev. A}
\textbf{87}, 022308 (2013).

\bibitem {endtoend}Saltzer, J. H., Reed, D. P. \& Clark, D. D. End-to-end
arguments in system design. \textit{Proceedings of the Second
International Conference on Distributed Computing Systems} (Paris,
France, April 8-10, 1981).

\bibitem {Baran}Baran, P. On distributed communications networks. \textit{IEEE
Trans. Commun.} \textbf{12}, pp. 1--9 (1964).

\bibitem {BellFORMULA}Spedalieri, G., Ottaviani, C. \& Pirandola, S.
Covariance matrices under Bell-like detections. \textit{Open Syst.
Inf. Dyn.} \textbf{20}, 1350011 (2013).

\bibitem {NOTA2way}This kind of decoding based on the subtraction of a
reference variable can also be found in two-way quantum
cryptography~\cite{TwowayPROTOCOL}.

\bibitem {Renner}Renner, R. Symmetry of large physical systems implies
independence of subsystems. \textit{Nature Phys.} \textbf{3}, 645
(2007).

\bibitem {Renner2}Renner, R. \& Cirac, J. I. de Finetti representation theorem
for infinite-dimensional quantum systems and applications to
quantum cryptography. \textit{Phys. Rev. Lett.} \textbf{102},
110504 (2009).

\bibitem {Raul}Garc\'{\i}a-Patr\'{o}n, R. \& Cerf, N. J. Unconditional
optimality of gaussian attacks against continuous-variable quantum
key distribution. \textit{Phys. Rev. Lett.} \textbf{97}, 190503
(2006).

\bibitem {canATTACKS}Pirandola, S., Braunstein, S. L. \& Lloyd, S.
Characterization of collective Gaussian attacks and security of
coherent-state quantum cryptography. \textit{Phys. Rev. Lett.}
\textbf{101}, 200504 (2008).

\bibitem {Pirandola2009}Pirandola, S., Garc\'{\i}a-Patr\'{o}n, R. Braunstein,
S.~L. \& Lloyd, S. Direct and reverse secret-key capacities of a
quantum channel. \textit{Phys. Rev. Lett.} \textbf{102}, 050503
(2009).

\bibitem {TwomodePRA}Pirandola, S., Serafini, A. \& Lloyd, S. Correlation
matrices of two-mode bosonic systems. \textit{Phys. Rev. A
\textbf{79}}, 052327 (2009).

\bibitem {NJP2013}Pirandola, S. Entanglement reactivation in separable
environments. \textit{New J. Phys.} \textbf{15}, 113046 (2013).

\bibitem {decoder}It is clear that the behaviour of the rate would be
perfectly inverted if Bob were the encoder of the secret
information and Alice the decoder.

\bibitem {Excess}Note that these values are much higher than those typically
appearing in experiments of continuous-variable QKD. For instance,
$\varepsilon\lesssim0.008$ in Ref.~\cite{Grangier2}.

\bibitem {Jouguet}Jouguet, P., Kunz-Jacques, S., \& Leverrier, A.
Long-distance continuous-variable quantum key distribution with a
Gaussian modulation. \textit{Phys. Rev. A} \textbf{84}, 062317
(2011).

\bibitem {Weed2010}Weedbrook, C., Pirandola, S., Lloyd, S. \& Ralph, T.~C.
Quantum cryptography approaching the classical limit.
\textit{Phys. Rev. Lett.} \textbf{105}, 110501 (2010).

\bibitem {Weed2012}Weedbrook, C., Pirandola, S. \& Ralph, T.~C.
Continuous-variable quantum key distribution using thermal states.
\textit{Phys. Rev. A} \textbf{86}, 022318 (2012).

\bibitem {Weed2013}Weedbrook, C., Ottaviani, C., \& Pirandola, S. Two-way
quantum cryptography at different wavelengths. \textit{Phys. Rev.
A} \textbf{89}, 012309 (2014).
\end{thebibliography}

\begin{thebibliography}{99}
\bibitem{RMPapp} Weedbrook, C., Pirandola, S., Garcia-Patron, R., Cerf, N.
J., Ralph, T. C., Shapiro, J. H. \& Lloyd, S. Gaussian quantum
information. \textit{Rev. Mod. Phys.} \textbf{84}, 621 (2012).

\bibitem{Raulapp} Garc\'{\i}a-Patr\'{o}n, R. \& Cerf, N. J. Unconditional
optimality of gaussian attacks against continuous-variable quantum
key distribution. \textit{Phys. Rev. Lett.}, \textbf{97}, 190503
(2006).

\bibitem{GaussSWAP} Pirandola, S., Vitali, D., Tombesi, P. \& Lloyd, S.
Macroscopic entanglement by entanglement swapping. \textit{Phys.
Rev. Lett.} \textbf{97}, 150403 (2006).

\bibitem{Pincibono} Picinbono, B. Second-order complex random vectors and
normal distributions. \textit{IEEE Trans. Sig. Proc.} \textbf{44},
2637--2640 (1996).

\bibitem{BellFORMULAapp} Spedalieri, G., Ottaviani, C. \& Pirandola, S.
Covariance matrices under Bell-like detections. \textit{Open Syst.
Inf. Dyn.} \textbf{20}, 1350011 (2013).

\bibitem{TwomodePRAapp} Pirandola, S., Serafini, A. \& Lloyd, S. Correlation
matrices of two-mode bosonic systems. \textit{Phys. Rev. A
\textbf{79}}, 052327 (2009).

\bibitem{NJP2013app} Pirandola, S. Entanglement reactivation in separable
environments. \textit{New J. Phys.} \textbf{15}, 113046 (2013).

\bibitem{NoSwitchapp} Weedbrook, C., Lance, A. M., Bowen, W. P., Symul, T.,
Ralph, T. C. \& Lam, P. K. Quantum cryptography without switching. \textit{%
Phys. Rev. Lett.} \textbf{93}, 170504 (2004).

\bibitem{TwowayPROTOCOLapp} Pirandola, S., Mancini, S., Lloyd, S. \&
Braunstein, S. L. Continuous-variable quantum cryptography with
two-way quantum communication. \textit{Nature Phys.} \textbf{4},
726 (2008).

\bibitem{SidePRLapp} Braunstein, S. L. \& Pirandola, S. Side-channel-free
quantum key distribution. \textit{Phys. Rev. Lett.} \textbf{108},
130502 (2012).

\bibitem{UlrikPRL} Niset, J., Ac\'{\i}n, A., Andersen, U. L., Cerf, N. J.,
Garc\'{\i}a-Patr\'{o}n, R., Navascu\'{e}s, M., \& Sabuncu, M.
Superiority of entangled measurements over all local strategies
for the estimation of product coherent states. \textit{Phys. Rev.
Lett.} \textbf{98}, 260404 (2007).

\bibitem{Grangier2app} Jouguet, P., Kunz-Jacques, S., Leverrier, A.,
Grangier, P. \& Diamanti, E. Experimental demonstration of
long-distance continuous-variable quantum key distribution.
\textit{Nature Photonics} \textbf{7}, 378--381 (2013).

\bibitem{Jouguetapp} Jouguet, P., Kunz-Jacques, S., \& Leverrier, A.
Long-distance continuous-variable quantum key distribution with a
Gaussian modulation. \textit{Phys. Rev. A} \textbf{84}, 062317
(2011).
\end{thebibliography}
\end{document}